\documentclass[10pt,journal,twoside]{IEEEtran}
\usepackage{multirow}
\usepackage{amssymb}
\usepackage{graphicx}
\usepackage{epsfig}
\usepackage{epstopdf}
\usepackage{caption}
\usepackage{amsmath}
\usepackage{amsthm}
\usepackage{array}
\usepackage{latexsym,bm}
\usepackage{color}
\usepackage{makecell}
\usepackage{subfigure}
\usepackage{longtable}
\usepackage{accents}
\usepackage{cite}
\usepackage{enumerate}
\usepackage{arydshln}

\usepackage[lined,ruled,linesnumbered]{algorithm2e}

\usepackage{stfloats}
\usepackage{diagbox}
\usepackage[english]{babel}
\usepackage[colorlinks=true, allcolors=blue]{hyperref}

\title{Design of A New Multiple-Chirp-Rate Index Modulation for LoRa Networks
\thanks{Xiaobin Zhu and Guofa Cai are with the School of Information Engineering, Guangdong University of Technology, Guangzhou 510006, China (e-mail: 2112303109@mail2.gdut.edu.cn; caiguofa2006@gdut.edu.cn)}
\thanks{Minling Zhang is with Guangzhou Information Technology Vocational School, Guangzhou 510091, China (e-mail: 2112103025@mail2.gdut.edu.cn)}
\thanks{Jiguang He is with Great Bay University, Dongguan 523000, China and Great Bay Institute for Advanced Study (GBIAS), Dongguan 523000, China (e-mail: jiguang.he@gbu.edu.cn)}
\thanks{Georges Kaddoum is with the Electrical Engineering Department, \'Ecole de Technologie Sup\'erieure, University of Qu\'ebec Montreal, QC H2L 2C4, Canada, and also with the Artificial Intelligence \& Cyber Systems Research Center, Lebanese American University (e-mail: georges.kaddoum@etsmtl.ca)}
}
\author{Xiaobin Zhu, Minling Zhang, Guofa Cai, \IEEEmembership{Senior Member}, \IEEEmembership{IEEE}, Jiguang~He, \IEEEmembership{Senior Member}, \IEEEmembership{IEEE}, Georges Kaddoum, \IEEEmembership{Senior Member}, \IEEEmembership{IEEE}}

\begin{document}
	\maketitle
	\begin{abstract}
	
	We propose a multiple chirp rate index modulation (MCR-IM) system based on Zadoff-Chu (ZC) sequences that overcomes the problems of low transmission rate and large-scale access in classical LoRa networks.
	We demonstrate the extremely low cross-correlation of MCR-IM signals across different spread factors, showing that the proposed MCR-IM system also inherits the characteristics of ZC sequences modulation.
	Moreover, we derive an approximate closed-form expression for the bit-error-rate (BER) of the proposed MCR-IM system over Nakagami-${m}$ fading channels.
	Simulation results confirm the accuracy of the derived closed-form expression and demonstrate that the MCR-IM system achieves higher levels of spectral efficiency (SE) compared to existing systems.
	In this context, assigning multiple chirp rates to each user results in a reduction in the number of parallel channels.
	To mitigate this issue, we propose a peak detection based successive interference cancellation (PD-SIC) algorithm to accommodate more users.
	Compared to orthogonal scatter chirp spreading spectrum system that names OrthoRa, the MCR-IM system with PD-SIC algorithm achieves lower BER levels.
	For a similar number of collision signals, the throughput of the MCR-IM system is enhanced by ${16\%}$ to ${21\%}$.
	Owing to these advantages, the proposed MCR-IM is well suited for large-scale, high-rate LoRa network applications.

	\end{abstract}


	\section{Introduction}
	Low-power wide-area networks (LPWAN) are critical for the current and future development of the Internet of Things (IoT), enabling low-power and long-distance communication for IoT devices.
	Long-range wide-area network (LoRaWAN) with LoRa (long-range) modulation has been widely investigated and found as one of the most promising LPWAN technologies.
	LoRaWAN operates in the unlicensed frequency band and typically adopts a star-of-stars topology, with hundreds or thousands of LoRa nodes connected to a single LoRa gateway.
	As a result, LoRaWAN has broad application prospects in intelligent transportation systems, smart cities, unmanned aerial vehicle communications and other fields \cite{9006936,9772404,9530545,10436105,10531262,10778617,10295375,10247030,10224843,10882982}.
	The main advantages of LoRa modulation are its high receiving sensitivity and long communication range.
	By adjusting the spreading factor (SF), LoRa can achieve multiple levels of spectral efficiency (SE) levels.
	Each LoRa symbol is generated by periodically shifting the frequency of a base chirp symbol.
	The information is modulated into the initial frequency using chirp spread spectrum (CSS), and the signal energy is concentrated into the corresponding frequency after dechirp operation \cite{8067462,8723130,9487489}.
	Due to these unique properties and superior performance, LoRa modulation has become a commonly used technology for LPWAN applications.
	
	Despite the significant potential of LoRa modulation, it still faces challenges in practical deployment \cite{9642432} \cite{10871926}.
	LoRaWAN uses the basic ALOHA protocol, which implies that the users transmit signals without sensing the channel status.
	Traditional LoRa enables parallel transmission only for users with different spreading factors.
	When two or more users with the same SF transmit information simultaneously, a collision occurs, preventing the gateway from demodulating the information correctly \cite{8903531} \cite{8678478}.
	Such colliding packets are then retransmitted after a random back-off time, which further exacerbates the collision problem in LoRa networks.
	To mitigate the packet collisions problem in LoRaWAN, particularly in large-scale device connectivity scenarios, significant research works have emerged.
	In \cite{9131834}\cite{10.1145/3098822.3098845}, different transmit nodes were assigned signals based on their carrier frequency offsets, thereby enabling the separation of colliding signals.
	In \cite{8888038} \cite{10.1145/3452296.3472931}, collisions are mitigated though interference reconstruction in time domain and frequency domain.
	In \cite{9664328} \cite{9646439}, subtle inter-packet time offsets were exploited to decompose concurrent transmissions and achieve a low signal-to-noise ratio (SNR) LoRa collision decoding.
	In \cite{10436392,10024800,9178997}, spectrograms were used to track signal continuity, find symbol boundaries to distinguish packets, and separate colluding symbols.
	These methods are often computationally expensive as they decode multiple signals simultaneously, rendering their implementation challenging \cite{9945838} \cite{9712392}.
	Another approach for solving the collision problem is to prevent signal interference through orthogonality.
	In \cite{9158432}, FlipLoRa, which leveraged the orthogonality between up-chirp and down-chirp modulation signals to reduce interference peaks in the frequency domain, was introduced to disentangle LoRa collisions.
	In \cite{10228983}, the orthogonal scatter chirp spreading spectrum system that names OrthoRa was proposed, which significantly enhances the concurrency for LPWAN transmission through a discrete orthogonal coding mechanism.
	Exploiting the quasi-orthogonality waveforms among user to mitigate collisions is a promising approach.
	However, the intricate interleaver design required at the transmitter for OrthoRa deployment presents a challenge for low-cost, large-scale applications.
	To address this issue, researchers have explored the use of Zadoff-Chu (ZC) sequences instead of traditional LoRa sequences for LPWAN \cite{9741074}.
	In \cite{10360184}, the authors validated the feasibility of replacing LoRa sequences with ZC sequences through simulations.
	By adjusting the different parameters, the ZC sequences exhibits a large number of quasi-orthogonal waveforms, surpassing by far the number generated by the LoRa sequences.
	This can potentially significantly increase the number of parallel channels \cite{1054840} \cite{1057786}.
	As a result, the ZC sequences are expected to replace LoRa sequences as the preferred choice for large-scale LPWAN in the future.
	
	Although the ZC sequences offer significant for large-scale IoT access, it suffers from a low transmission rate, similar to the LoRa sequences.
	This limitation poses a challenge for the practical deployment of LoRa, particularly in IoT applications with high transmission rate requirements.
	To improve the transmission rate of LoRa modulation, many variants have been proposed.
	In \cite{8607020} \cite{9207749}, interleaved chirp spreading LoRa (ICS-LoRa) was proposed sending an extra bit of data by adding an interleaver to LoRa.
	In \cite{9542955} \cite{9348094}, I/Q components were used in LoRa modulation to encode additional information.
	In \cite{8746470} \cite{9663547}, phase-shift keying (PSK) modulation was added to LoRa, i.e., PSK-LoRa, in order to send extra information.
	In \cite{9123393,9709288,9693529}, slope-shift keying LoRa (SSK-LoRa) and time domain multiplexed LoRa (TDM-LoRa) were proposed to encode one additional bit of information by utilizing the up-chirp and down-chirp.
	In \cite{9446168}, discrete chirp rate keying LoRa (DCRK-LoRa) was presented to transmit more information by expanding the selection of chirp rates.
	However, the orthogonality constraint of LoRa sequences limits the range of selectable chirp rates for DCRK-LoRa.
	In \cite{8937880,9672170,9828505,9659806}, dual orthogonal LoRa (DO-LoRa) and group-based CSS (GCSS) were proposed to achieve independent transmission of multiple signals by grouping frequencies.
	In \cite{9435804}, frequency-shift CSS with index modulation (FSCSS-IM) using multiple activated frequencies was proposed, representing a typical application of index modulation in CSS techniques.
	Those systems with phase or I/Q components to modulate information typically necessitate coherent demodulation \cite{10391276}.
	This increases receiver complexity while reducing system generality.
	The LoRa sequences also limit the use of signals with different chirp rates to transmit information.
	This is because the orthogonality of the LoRa sequences significantly deteriorates at different chirp rates, whereas ZC sequences can compensate for this orthogonality shortcoming.
	
	Inspired by the above mentioned motivations, in this paper, we propose a multiple chirp rate IM (MCR-IM) system utilizing ZC sequences.
	Leveraging the extensive set of orthogonal waveforms provided by ZC sequences, the MCR-IM system can assign signals with different chirp rate ranges to different users.
	This enables simultaneous transmission while ensuring correct demodulation at the receiver.
	Moreover, by extending the index range to multiple chirp rates, the MCR-IM system can achieve a higher upper bound on transmission rate.
	The contributions of this paper are summarized as follows:
	\begin{itemize}
		\item
		We propose an MCR-IM system based on ZC sequences.
		By expanding the index range of the chirp rate, the MCR-IM system achieves a higher transmission rate than the existing LoRa system.
		With the same number of indexes, the proposed system can encode more information bits.
		Simulation results reveal that the MCR-IM system achieves a higher SE than existing systems.
		\item
		We show that MCR-IM signals with different SFs are quasi-orthogonal.
		Moreover, we theoretically analyze the bit-error-rate (BER) performance of the MCR-IM system over Nakagami-${m}$ fading channels and derive a corresponding closed-form expression.
		The theoretical results are consistent with the numerical results.
		\item
		To address the multi-user detection challenge, we propose a peak detection based successive interference cancellation (PD-SIC) algorithm to improve the decoding performance of the MCR-IM system.
		Simulation results show that the BER of the MCR-IM system with PD-SIC algorithm reaches ${10^{-5}}$ and achieves a higher throughput than the OrthoRa system.
	\end{itemize}

	The remainder of this paper is organized as follows.
	Section II introduces the proposed MCR-IM system, detailing the design of both the transmitter and receiver.
	Section III analyzes the BER performance of the MCR-IM system with non-coherent demodulation over Nakagami-${m}$ fading channels.
	Section IV discusses the signal collision problem and proposes the PD-SIC algorithm to optimize the system performance.
	Section V presents the simulation results and corresponding discussions.
	Finally, Section VI concludes the work.
	
	\textit{Notations}:
	Bold lowercase letters denotes vectors (e.g., ${\bf{a}}$), while bold capital letters represent matrices (e.g., ${\bf{A}}$).
	${{\bf{a}}\left( k \right)}$ denotes the ${k}$-th element of the vector ${\bf{a}}$.
	${\left| {\bf{A}} \right|}$ returns the absolute value of each element in matrix ${{\bf{A}}}$.
	The operators ${\left(  \cdot  \right)^ * }$ and $\left(  \cdot  \right)^\mathsf{T}$ denotes conjugate of a complex number and the matrix or vector transpose, respectively.
	Blackboard bold letters denote the sets (e.g., ${\mathbb{A}}$).
	${\left| {\mathbb{A}} \right|}$ denotes the cardinality of the set ${\mathbb{A}}$.
	${\mathbb{A}\left( k \right)}$ denotes the ${k}$-th element of the set ${\mathbb{A}}$.
	${\left\lfloor a \right\rfloor }$ denotes the floor function, rounding ${a}$ down to the nearest integer.
	${\binom{a}{b}}$ denotes the combination number calculation ${\frac{{a!}}{{b!\left( {a - b} \right)!}}}$, when ${a < b}$, ${\binom{a}{b}=0}$.
	${a!}$ stands for factorial operation.
	${\bmod \left( {a,b} \right)}$ returns the remainder when ${a}$ is divided by ${b}$.
	${{\rm{sort}}\left(  \cdot  \right)}$ stands for the ascending sort operation.
	${{\cal C}{\cal N}\left( {a,b} \right)}$ denotes the complex Gaussian distribution with mean ${a}$ and variance ${b}$.
	% Finally, $\Re\{\cdot\}$ returns the real part of its complex argument.
	
	\section{System Model}
	
	\begin{figure*}
		\centering
		\includegraphics[scale=0.7]{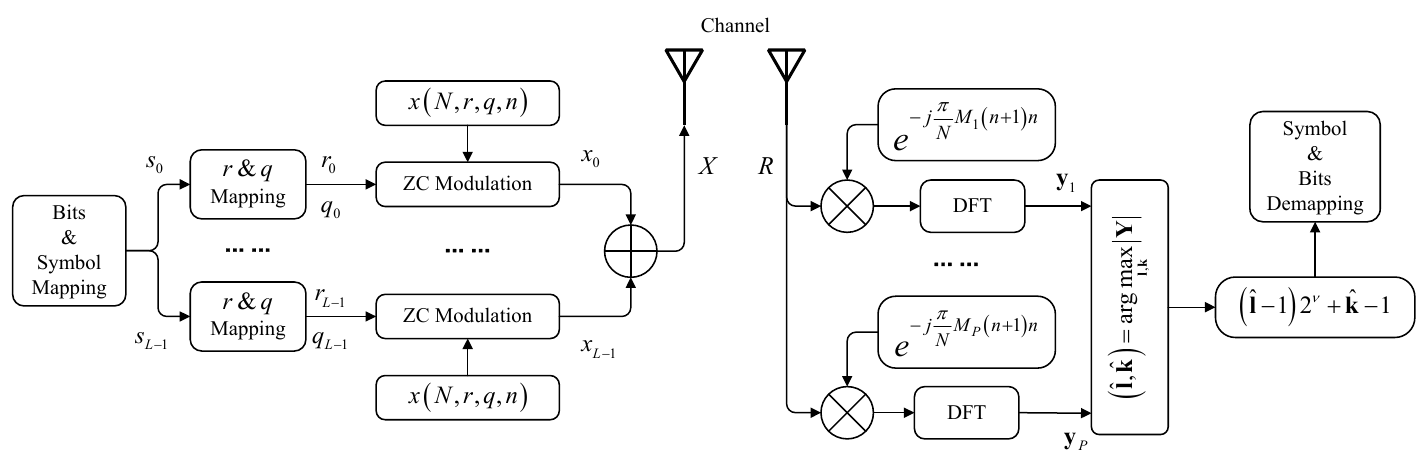}
		\caption{Block diagram of the proposed MCR-IM system.}
		\label{fig01}
	\end{figure*}
	
	In this section, the proposed MCR-IM system, which includes both the transmitter and receiver, as shown in Fig.~\ref{fig01}, is introduced.

	\subsection{ZC Sequences Modulation}
	
	\begin{table}[t]
		\renewcommand{\arraystretch}{2}		% 调整表格高度
		\setlength{\tabcolsep}{7pt}			% 调整表格宽度
		\centering							% 居中
		\caption{One-to-one correspondence between ${\nu}$ and ${N}$ in the ZC sequences.}
		\label{table:Table SF N}
		\begin{tabular}{c|c|c|c|c|c|c|c}	% 标注多少列，哪些列不要竖线
			\Xhline{1pt}
			${\nu}$ & 6  & 7   & 8   & 9   & 10   & 11   & 12   \\ \hline
			${N}$  & 67 & 131 & 257 & 521 & 1031 & 2053 & 4099 \\
			\Xhline{1pt}
		\end{tabular}
	\end{table}

	According to \cite{10360184}, the ZC sequences can be expressed as
	\begin{align}
		x\left( {N,r,q,n} \right) = \frac{1}{{\sqrt N }}\exp \left( {j\frac{\pi }{N}r\left( {n + 1 + 2q} \right)n} \right),
	\end{align}
	where ${N}$ is a prime number, ${r \in \left\{ {1,2, \cdots ,N - 1} \right\}}$ denotes the root or chirp rate, ${q \in \left\{ {0,1, \cdots ,N - 1} \right\}}$ is the offset and ${n = 0,1, \cdots, N - 1}$.
	The ZC sequences exhibit a lower cross-correlation than traditional LoRa sequences, particularly when the signals have different chirp rates.
	The proof is provided in Appendix \ref{proff LoRa} and Appendix \ref{proff A}.
	
	To facilitate comparison with the LoRa sequences, we define the SF of the ZC sequences as ${\nu  = \left\lfloor {{{\log }_2}N} \right\rfloor }$.
	According to \cite{10360184}, truncating the ZC sequences to match the length of the LoRa sequences is feasible.
	The truncated ZC sequences retains its advantageous properties.
	Unless otherwise specified, the one-to-one correspondence between ${\nu}$ and ${N}$ in this paper is shown in Table \ref{table:Table SF N}.
	
	For demodulation, the conjugate ZC sequences used for the dechirp operation can be given by
	\begin{align}
		{x^ * }\left( {N,r',0,n} \right) = \frac{1}{{\sqrt N }}\exp \left( { - j\frac{\pi }{N}r'\left( {n + 1} \right)n} \right).
	\end{align}

	The demodulation result is obtained using the discrete fourier transform (DFT), given by
	\begin{align}
		y\left( k \right) &= \left| {{\rm{DFT}}\left( {x\left( {N,r,q,n} \right){x^ * }\left( {N,r',0,n} \right)} \right)} \right|\nonumber\\
		&= \frac{1}{N}\left| {\sum\limits_{n = 0}^{N - 1} {{e^{j\frac{{\pi n}}{N}\left( {\left( {r - r'} \right)\left( {n + 1} \right) + 2\left( {rq - k} \right)} \right)}}} } \right|,
	\end{align}
	where ${k = 0,1, \cdots N - 1}$.
	When ${r = {r'}}$, we get
	\begin{align}
		y\left( k \right) = \left\{ {\begin{array}{*{20}{l}}
				{1\qquad k = \,\bmod \,\left( {rq,N} \right),}\\
				{0\qquad {\rm{other}}{\rm{.}}}
		\end{array}} \right.
	\end{align}
	
	When ${r \ne {r'}}$, ${y\left( k \right) = \frac{1}{{\sqrt N }}}$.
	This is one of the key features of the ZC sequences, where its energy is uniformly distributed across all frequencies when the chirp rates mismatch.
	When ZC sequences with different chirp rates are indexed, the peak frequency can be demodulated using the conjugate sequences of the corresponding chirp rate, while non-matching sequences are dispersed across frequencies as interference.
	In addition, it can be observed that the impact of such interference diminishes as ${N}$ increases.
	
	\subsection{MCR-IM Transmitter}

	As shown in Fig.~\ref{fig01}, at the transmitter, the chirp rate ${r}$ can be selected within the range ${\left[ {{M_1},{M_P}} \right]}$, yielding a total of ${P}$ available chirp rates.
	We assume that the number of indexed signals is ${L}$.
	Since signals with different chirp rates exhibit a low cross-correlation, they can be considered relatively independent.
	
	For ease of analysis, we only select the first ${2^{\nu}}$ frequencies for activation within each chirp rate, resulting in a total of ${2^{\nu}P}$ frequencies available for selection.
	This can be represented as a ${P \times 2^{\nu}}$ matrix ${{\bf{\Theta }}}$, where each row corresponds to a different chirp rate and each column corresponds to a different frequency.
	The system selects ${L}$ elements from the matrix ${{\bf{\Theta }}}$ to transmit data as a symbol ${{\bf{s}}}$, where ${{\bf{s}} = \left[ {{s_0},{s_1}, \cdots ,{s_{L - 1}}} \right]}$, ${{s_i} \in \left\{ {0,1, \cdots 2^{\nu}P - 1} \right\}}$, ${i = 0,1, \cdots ,L - 1}$, and ${{s_0} < {s_1} <  \cdots  < {s_{L - 1}}}$.
	Hence, the number of transmitted bits in a symbol can be computed as ${{{\eta _b}} = \left\lfloor \log_2\binom{2^{\nu}P}{L} \right\rfloor}$.

	The challenge lies in mapping ${{\eta _b}}$ binary bits into ${{\bf{s}}}$, given the vast number of possible combinations for ${{\bf{s}}}$, which can be as large as ${\binom{2^{\nu}P}{L}}$.
	Due to the large memory requirements of lookup tables (LUTs), the combinatorial method \cite{6587554} is adopted here instead.
	First, we convert ${{\eta _b}}$ binary number to a decimal numbers ${D}$.
	For any natural number ${D \in \left\{ {0,1,2, \cdots ,\binom{2^{\nu}P}{L}- 1} \right\}}$, the combinatorial method maps ${D}$ to a strictly monotonically decreasing sequences of length ${L}$, which is expressed as
	\begin{align}
		D = \binom{s_{L-1}}{L} + \binom{s_{L-2}}{L-1} +  \cdots  + \binom{s_1}{2} + \binom{s_0}{1}.
	\end{align}

	Subsequently, the generation method of ${{\bf{s}}}$ can be described as:
	1) select the largest ${s_{L-1}}$ such that ${\binom{s_{L-1}}{L} \le D}$;
	2) select the largest ${s_{L-2}}$ such that ${\binom{s_{L-2}}{L-1} \le D - \binom{s_{L-1}}{L}}$; and so on; until all ${L}$ elements of the ${{\bf{s}}}$ are obtained.
	For instance, when ${L=3}$, ${P=2}$, and ${2^{\nu}=8}$, there are a total number of ${\binom{2 \times 8}{3} = 560}$ combinations for which the corresponding sequences ${{\bf{s}}}$ can be obtained as
	\begin{equation}
		\begin{aligned}
		&0 = \binom{2}{3} + \binom{1}{2} + \binom{0}{1} \to {\bf{s}} = \left[ {0,1,2} \right], \nonumber\\
		&1 = \binom{3}{3} + \binom{1}{2} + \binom{0}{1} \to {\bf{s}} = \left[ {0,1,3} \right], \nonumber\\
		&\vdots \nonumber\\
		&558 = \binom{15}{3} + \binom{14}{2} + \binom{12}{1} \to {\bf{s}} = \left[ {12,14,15} \right], \nonumber\\
		&559 = \binom{15}{3} + \binom{14}{2} + \binom{13}{1} \to {\bf{s}} = \left[ {13,14,15} \right]. \nonumber\\
		\end{aligned}
	\end{equation}
	Thus, we obtain the required ${L}$ index elements.

	Additionally, each element ${{s_i}}$ has a corresponding ${{q_i}}$ and ${{r_i}}$, where
	\begin{align}
		\label{modu 8}
		{{r_i} = \left\lfloor {\frac{{{s_i}}}{{{2^{\nu}}}}} \right\rfloor  + {M_1}},
	\end{align}
	\begin{align}
		{{s_i} - \left\lfloor {\frac{{{s_i}}}{{{2^{\nu}}}}} \right\rfloor {2^{\nu}} = \bmod \left( {{r_i}{q_i},N} \right)}.
	\end{align}
	We therefore obtain ${{\bf{r}} = \left[ {{r_0},{r_1}, \cdots ,{r_{L - 1}}} \right]}$ and ${{\bf{q}} = \left[ {{q_0},{q_1}, \cdots ,{q_{L - 1}}} \right]}$.
	Hence, the signal ${{x_i}}$ corresponding to each symbol ${{s_i}}$ is given by
	\begin{align}
		\label{modu 10}
		{x_i}\left( {N,{r_i},{q_i},n} \right) = \exp \left( {j\frac{\pi }{N}{r_i}\left( {n + 1 + 2{q_i}} \right)n} \right).
	\end{align}

	After stacking ${L}$ signals together, the transmitted signal of the proposed MCR-IM system is obtained as
	\begin{align}
		\label{modu 11}
		X\left( n \right) = \sqrt {\frac{{{E_{\rm{s}}}}}{{NL}}} \sum\limits_{i = 0}^{L - 1} {\exp \left( {j\frac{\pi }{N}{r_i}\left( {n + 1 + 2{q_i}} \right)n} \right)},
	\end{align}
	where ${{{E_{\rm{s}}}}}$ represents the symbol energy, and the energy per bit is given by ${{E_{\rm{b}}} = \frac{{{E_{\rm{s}}}}}{{{\eta _b}}}}$.
	The proof of the orthogonality of MCR-IM signals across different SFs is provided in Appendix \ref{proff C}.
	Moreover, the MCR-IM inherits the characteristics of the ZC sequences based modulation and LoRa modulation.

	\subsection{MCR-IM Receiver}
	
	\begin{figure}
		\centering
		\includegraphics[scale=0.32]{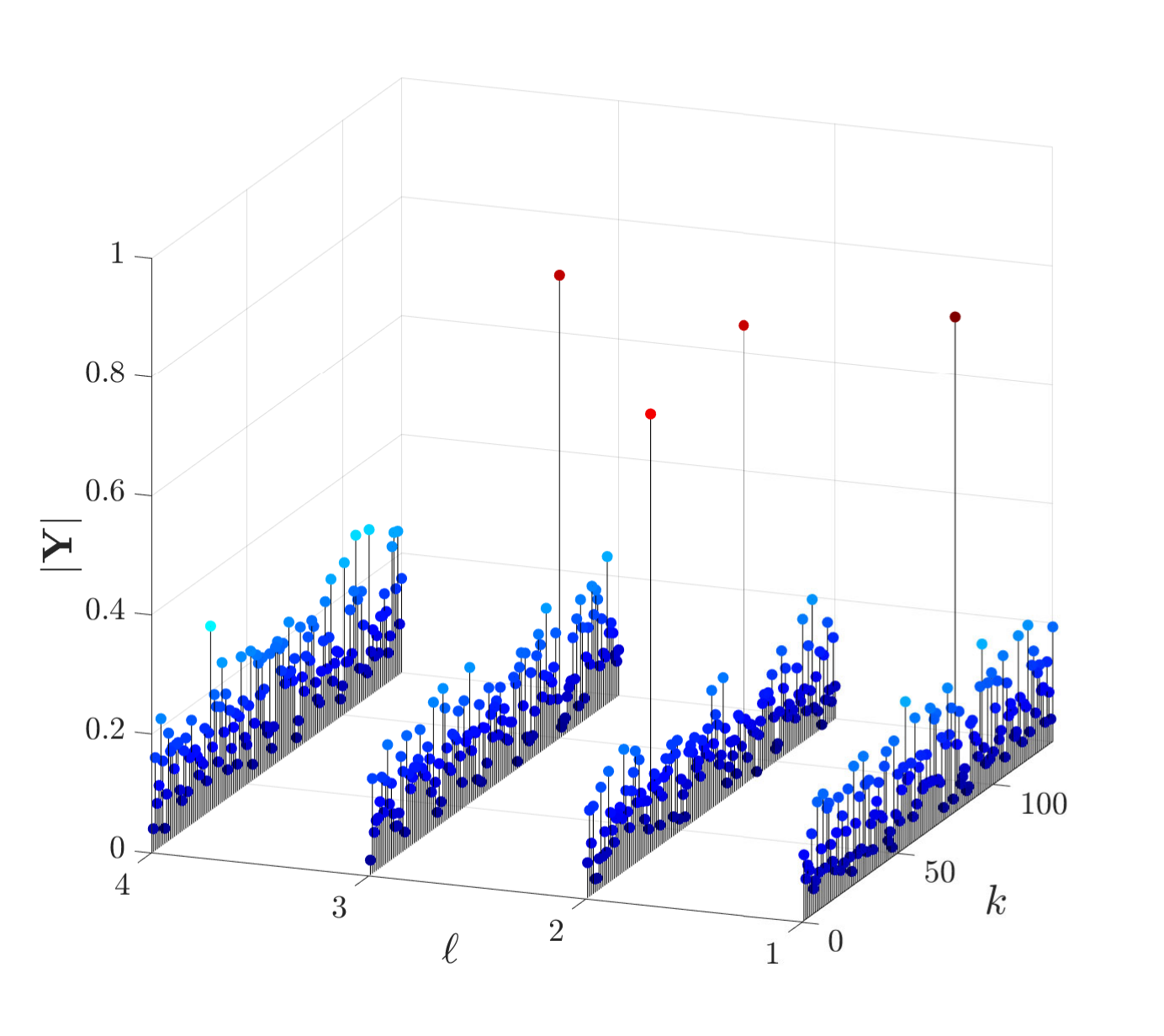}
		\caption{${\left| {\bf{Y}} \right|}$ under Nakagami-${m}$ fading channels, where ${\nu = 7}$, ${N=131}$, ${\left[ {{M_1},{M_P}} \right] = \left[ {1,4} \right]}$, ${L = 4}$, ${{{E_{\rm{s}}}} = 1}$, ${{\bf{s}} = \left[ {79,210,420,483} \right]}$, ${\frac{{{E_{\rm{b}}}}}{{{N_0}}} = 10}$ dB and ${m=3}$.}
		\label{fig02}
	\end{figure}
	
	After the transmitted signal ${X\left( n \right)}$ passes through a Nakagami-${m}$ fading channel, the signal ${R\left( n \right)}$ at the receiver can be expressed as
	\begin{align}
		R\left( n \right) = hX\left( n \right) + \phi \left( n \right),
	\end{align}
	where ${\phi \left( n \right)}$ denotes the complex additive white Gaussian noise (AWGN) follows ${{\cal C}{\cal N}\left( {0,\frac{{{N_0}}}{2}} \right)}$.
	${h}$ represents the complex channel gain whose envelope ${\left| h \right|}$ follows a Nakagami-${m}$ distribution with shape parameter ${m}$.
	The probability density function (PDF) of ${\left| h \right|}$ is expressed as
	\begin{align}
		{f_m}\left( x \right) = \left\{ {\begin{array}{*{20}{c}}
				{\frac{2}{{\Gamma \left( m \right)}}{{\left( {\frac{m}{\Omega }} \right)}^m}{x^{2m - 1}}\exp \left( { - \frac{{m{x^2}}}{\Omega }} \right),x > 0,}\\
				{0,\qquad {\rm{other}},}
		\end{array}} \right.
	\end{align}
	where ${\Omega  = 1}$.
	At the receiver, conjugate chirp signals with different chirp rates are used for demodulation.
	For different chirp rates ${\left\{ {{M_1}, \cdots ,{M_P}} \right\}}$, we obtain different demodulation results as
	\begin{align}
		\label{demo_01}
		{{\bf{y}}_\ell }\left( k \right) &= {\rm{DFT}}\left( {R\left( n \right)\frac{1}{{\sqrt N }}{e^{\left( { - j\frac{\pi }{N}{M_\ell }\left( {n + 1} \right)n} \right)}}} \right)\nonumber\\
		&= \frac{h}{{\sqrt N }}\sum\limits_{n = 0}^{N - 1} {X\left( n \right){e^{\left( { - j\frac{\pi }{N}{M_\ell }\left( {n + 1} \right)n} \right)}}{e^{\left( { - j\frac{{2\pi }}{N}kn} \right)}}}  + {\Phi _\ell }\left( k \right)\nonumber\\
		&= h{{\bf{u}}_\ell }\left( k \right) + {\Phi _\ell }\left( k \right),
	\end{align}
	where ${{{\bf{u}}_\ell }\left( k \right)}$ represents the signal component after the DFT, ${{{\Phi }_\ell }\left( k \right)}$ represents the noise and follows ${{\cal C}{\cal N}\left( {0,\frac{{{N_0}}}{2}} \right)}$, ${\ell  \in \left\{ {1,2, \cdots P} \right\}}$ and ${k = 0,1, \cdots {2^{\nu}} - 1}$.
	The ${P \times 2^{\nu}}$ matrix ${{\bf{Y}}}$ is defined as
	\begin{align}
		\label{demo_02}
		{\bf{Y}} = {\left[ {{{\bf{y}}_1};{{\bf{y}}_2}; \cdots ;{{\bf{y}}_P}} \right]^{\mathsf{T}}}.
	\end{align}

	In the case of non-coherent demodulation, we take the absolute value of each element in ${{\bf{Y}}}$.
	For instance, Fig.~\ref{fig02} illustrates ${\left| {\bf{Y}} \right|}$ obtained under Nakagami-${m}$ fading channels, where ${\frac{{{E_{\rm{b}}}}}{{{N_0}}}}$ denotes the SNR.
	Next, ${L}$ maximum points in ${\left| {\bf{Y}} \right|}$ are detected and their corresponding row indexes ${{\bf{\hat l}} = \left[ {{{\hat \ell }_0},{{\hat \ell }_1}, \cdots ,{{\hat \ell }_{L - 1}}} \right]}$ and columns indexes ${{\bf{\hat k}} = \left[ {{{\hat k}_0},{{\hat k}_1}, \cdots ,{{\hat k}_{L - 1}}} \right]}$ are determined as
	\begin{align}
		\label{demo 23}
		\left( {{\bf{\hat l}},{\bf{\hat k}}} \right) = \arg \mathop {\max }\limits_{{\bf{l}},{\bf{k}}} \left| {\bf{Y}} \right|,
	\end{align}
	where ${{{\hat \ell }_i} \in \left\{ {1,2, \cdots ,P} \right\}}$, ${{{\hat k}_i} \in \left\{ {1,2, \cdots ,{2^{\nu}}} \right\}}$, and ${i = 0,1, \cdots ,L - 1}$.
	Therefore, one can obtain
	\begin{align}
		\label{demo_24}
		{{\hat s}_i} = \left( {{{\hat \ell }_i} - 1} \right){2^{\nu}} + {{\hat k}_i} - 1.
	\end{align}
	
	After sorting, the estimated symbol ${{{\bf{\tilde s}}}}$ is obtained as
	\begin{align}
		\label{demo 25}
		{\bf{\tilde s}} = [{{\tilde s}_0},{{\tilde s}_1}, \cdots ,{{\tilde s}_{L - 1}}] = {\rm{sort}}\left( {{{\hat s}_0},{{\hat s}_1}, \cdots ,{{\hat s}_{L - 1}}} \right),
	\end{align}
	where ${{{\tilde s}_0} < {{\tilde s}_1} <  \cdots  < {{\tilde s}_{L - 1}}}$.
	The combinatorial method is adopted to obtain ${{\tilde D}}$, computed as
	\begin{align}
		{\tilde D} = \binom{{{\tilde s}_{L - 1}}}{L} + \binom{{{\tilde s}_{L - 2}}}{L-1} +  \cdots  + \binom{{{\tilde s}_1}}{2} + \binom{{{\tilde s}_0}}{1}.
	\end{align}

	Finally, ${{\tilde D}}$ is converted into a binary number to complete the demodulation process.
	
	\section{Performance Analysis}
	In this section, we analyze the BER performance of the proposed MCR-IM system.
	
	For ease of analysis and description, we define the signal matrix ${{\bf{U}} = {\left[ {{{\bf{u}}_1},{{\bf{u}}_2}, \cdots ,{{\bf{u}}_P}} \right]^{\mathsf{T}}}}$ and noise matrix ${{\bf{\Phi }} = {\left[ {{{\Phi }_1};{{\Phi }_2}; \cdots ;{{\Phi }_P}} \right]^{\mathsf{T}}}}$. 	
	Hence, one obtains
	\begin{align}
		{\bf{Y}} = h{\bf{U}} + {\bf{\Phi }}.
	\end{align}
	
	In general, ${\left| {{{\bf{y}}_\ell }\left( k \right)} \right|}$ follows a Rician distribution with shape parameter ${{\kappa _\ell }\left( k \right) = \frac{{{{\left| h {{{\bf{u}}_\ell }\left( k \right)} \right|}^2}}}{{{N_0}}}}$.
	We divide the elements in matrix ${{\bf{Y}}}$ into three categories, i.e., the signal set ${{{\mathbb{A}}_{\rm{s}}}}$, the interference set ${{{\mathbb{A}}_{\rm{i}}}}$ and the noise set ${{{\mathbb{A}}_{\rm{n}}}}$.
	
	For a determined symbol ${{\bf{s}}}$, it can uniquely correspond to the matrix ${\bf{U}}$.
	If ${\left( {\ell  - 1} \right)2^{\nu} + k - 1 \in {\left\{ {{s_0},{s_1}, \cdots ,{s_{L - 1}}} \right\}}}$, this means that the ${{{\bf{y}}_\ell }\left( k \right)}$ is the peak point, then ${{{\bf{y}}_\ell }\left( k \right) \in {{\mathbb{A}}_{\rm{s}}}}$, otherwise ${{{\bf{y}}_\ell }\left( k \right) \in {{\mathbb{A}}_{\rm{i}}}}$ or ${{{\bf{y}}_\ell }\left( k \right) \in {{\mathbb{A}}_{\rm{n}}}}$.
	According to \cite{8607020}, because the ratio relationship between ${{{\bf{y}}_\ell }\left( k \right) \in {{\mathbb{A}}_{\rm{s}}}}$ and ${{{\bf{y}}_\ell }\left( k \right) \notin {{\mathbb{A}}_{\rm{s}}}}$ is adopted to determine the threshold, the influence of channel fading ${h}$ will be offset. Thus, the threshold of 1.3 dB  in \cite{8607020} can be also used to distinguish between the interference and noise sets. 
When ${{{\bf{y}}_\ell }\left( k \right) \notin {{\mathbb{A}}_{\rm{s}}}}$, if ${{\kappa _\ell }\left( k \right) \le 1.3}$ dB, then ${{{\bf{y}}_\ell }\left( k \right) \in {{\mathbb{A}}_{\rm{n}}}}$, otherwise ${{{\bf{y}}_\ell }\left( k \right) \in {{\mathbb{A}}_{\rm{i}}}}$.
	
	If ${{{{\bf{y}}}_\ell }\left( k \right) \in {{\mathbb{A}}_{\rm{s}}}}$ or ${{{\bf{y}}_\ell }\left( k \right) \in {{\mathbb{A}}_{\rm{i}}}}$, ${\left| {{{\bf{y}}_\ell }\left( k \right)} \right|}$ can be approximated as a Gaussian distribution with mean and variance given by
	\begin{align}
		\label{mu}
		{\mu _{\ell ,k}} = \sigma \sqrt {\frac{\pi }{2}} {}_1{F_1}\left( { - \frac{1}{2},1, - {\kappa _\ell }\left( k \right)} \right),
	\end{align}
	\begin{align}
		\label{sigma}
		\sigma _{\ell ,k}^2 = {N_0} + {\left| {h{{{\bf{u}}}_\ell }\left( k \right)} \right|^2} - \frac{{\pi {N_0}}}{4}{}_1F_1^2\left( { - \frac{1}{2},1, - {\kappa _\ell }\left( k \right)} \right),
	\end{align}
	where ${_1{F_1}\left( {a,b;x} \right) = \sum\limits_{k = 0}^\infty  {\frac{{\Gamma \left( {a + k} \right)\Gamma \left( b \right){x^k}}}{{\Gamma \left( a \right)\Gamma \left( {b + k} \right)k!}}} }$ is the confluent hypergeometric function.
	
	For the case where ${{{\bf{y}}_\ell }\left( k \right) \in {{\mathbb{A}}_{\rm{n}}}}$, ${\left| {{{\bf{y}}_\ell }\left( k \right)} \right|}$ can be approximated by the Rayleigh distribution with scale parameter ${{\sigma _n}}$ satisfying ${2\sigma _n^2 = {\left| {h{{{\bf{u}}}_\ell }\left( k \right)} \right|^2} + {N_0}}$.
	According to \cite{8392707}, the variance of ${\left| {{{\bf{y}}_\ell }\left( k \right)} \right|}$ is very low, and thus the maximum magnitude of ${\left| {{{\bf{y}}_\ell }\left( k \right)} \right|}$ can be approximated as a constant, i.e.,
	\begin{align}
		{\max _{{{\mathbb{A}}_{\rm{n}}}}}\left| {{{\bf{y}}_\ell }\left( k \right)} \right| \approx \sqrt {\left( {\bar U_n^2 + {N_0}} \right){H_{\left| {{{\mathbb{A}}_{\rm{n}}}} \right|}}},
	\end{align}
	\begin{align}
		{{\bar U}_n} = \frac{{\left| h \right|}}{{\left| {{{\mathbb{A}}_{\rm{n}}}} \right|}}\sum\limits_{{{\mathbb{A}}_{\rm{n}}}} {\left| {{{\bf{u}}_\ell }\left( k \right)} \right|},
	\end{align}
	where ${{H_i} = \sum\limits_{n = 1}^i {\frac{1}{n}} }$, ${{{\bar U}_n}}$ denotes the average absolute value of ${{{{\bf{u}}_\ell }\left( k \right)}}$ in the set ${{{{\mathbb{A}}_{\rm{n}}}}}$ under Nakagami-${m}$ fading.
	
	The symbol error probability  consists of two parts, i.e., one affected by ${{{\mathbb{A}}_{\rm{i}}}}$ and the other by ${{{\mathbb{A}}_{\rm{n}}}}$.
	To ensure that all the elements in the signal set ${{{\mathbb{A}}_{\rm{s}}}}$ can be correctly detected, all ${\left| {{{\bf{y}}_\ell }\left( k \right)} \right|}$'s in ${{{\mathbb{A}}_{\rm{s}}}}$ have to be greater than ${\left| {{{\bf{y}}_\ell }\left( k \right)} \right|}$ in ${{{\mathbb{A}}_{\rm{i}}}}$.
	Its probability of correct detection is the product of all individual success probabilities ${{\Pr \left( {{{\mathbb{A}}_{\rm{s}}}\left( a \right) > {{\mathbb{A}}_{\rm{i}}}\left( b \right)} \right)} }$.
	Then, one can obtain the  symbol error probability of the part affected by ${{{\mathbb{A}}_{\rm{i}}}}$ as follows
	\begin{align}
		{P_{{{\mathbb{A}}_{\rm{i}}}}} &= 1 - \prod\limits_{a = 1}^{\left| {{{\mathbb{A}}_{\rm{s}}}} \right|} {\prod\limits_{b = 1}^{\left| {{{\mathbb{A}}_{\rm{i}}}} \right|} {\Pr \left( {{{\mathbb{A}}_{\rm{s}}}\left( a \right) > {{\mathbb{A}}_{\rm{i}}}\left( b \right)} \right)} } \nonumber \\
		&= 1 - \prod\limits_{a = 1}^{\left| {{{\mathbb{A}}_{\rm{s}}}} \right|} {\prod\limits_{b = 1}^{\left| {{{\mathbb{A}}_{\rm{i}}}} \right|} {Q\left( {\frac{{{\mu _{{{\mathbb{A}}_{\rm{i}}}\left( b \right)}} - {\mu _{{{\mathbb{A}}_{\rm{s}}}\left( a \right)}}}}{{\sqrt {\sigma _{{{\mathbb{A}}_{\rm{s}}}\left( a \right)}^2 + \sigma _{{{\mathbb{A}}_{\rm{i}}}\left( b \right)}^2} }}} \right)} },
	\end{align}
	${{{\mu _{{{\mathbb{A}}_{\rm{s}}}\left( a \right)}}}}$ and ${{\sigma _{{{\mathbb{A}}_{\rm{s}}}\left( a \right)}^2}}$ represent the mean and variance of the ${a}$-th element ${\left| {{{\bf{y}}_\ell }\left( k \right)} \right|}$ in the set ${{{{\mathbb{A}}_{\rm{s}}}}}$, and ${{{\mu _{{{\mathbb{A}}_{\rm{i}}}\left( b \right)}}}}$ and ${{\sigma _{{{\mathbb{A}}_{\rm{i}}}\left( b \right)}^2}}$ represent the mean and variance of the ${b}$-th element ${\left| {{{\bf{y}}_\ell }\left( k \right)} \right|}$ in the set ${{{{\mathbb{A}}_{\rm{i}}}}}$, respectively, which can be computed using (\ref{mu}) and (\ref{sigma}), respectively.
	
	For the effect of ${{{\mathbb{A}}_{\rm{n}}}}$, all ${\left| {{{\bf{y}}_\ell }\left( k \right)} \right|}$ in ${{{\mathbb{A}}_{\rm{s}}}}$ have to be greater than the constant ${\sqrt {\left( {\bar U_n^2 + {N_0}} \right){H_{\left| {{{\mathbb{A}}_{\rm{n}}}} \right|}}} }$.
	Similarly, one can obtain the  symbol error probability of the part affected by ${{{\mathbb{A}}_{\rm{n}}}}$ as follows
	\begin{align}
		{P_{{{\mathbb{A}}_{\rm{n}}}}} &= 1 - \prod\limits_{a = 1}^{\left| {{{\mathbb{A}}_{\rm{s}}}} \right|} {\Pr \left( {{{\mathbb{A}}_{\rm{s}}}\left( a \right) > \sqrt {\left( {\bar U_n^2 + {N_0}} \right){H_{\left| {{{\mathbb{A}}_{\rm{n}}}} \right|}}} } \right)} \nonumber \\
		&= 1 - \prod\limits_{a = 1}^{\left| {{{\mathbb{A}}_{\rm{s}}}} \right|} {Q\left( {\frac{{\sqrt {\left( {\bar U_n^2 + {N_0}} \right){H_{\left| {{{\mathbb{A}}_{\rm{n}}}} \right|}}}  - {\mu _{{{\mathbb{A}}_{\rm{s}}}\left( a \right)}}}}{{\sqrt {\sigma _{{{\mathbb{A}}_{\rm{s}}}\left( a \right)}^2} }}} \right)}.
	\end{align}

	%The proposed modulation can fall under the category of orthogonal signaling.
Hence, with a channel gain ${\left| h \right| = \alpha }$, the total BER can be expressed as
	\begin{align}
		{P_\alpha } = \frac{2^{\eta _b-1}}{2^{\eta _b}-1}\left( {1 - \left( {1 - {P_{{{\mathbb{A}}_{\rm{i}}}}}} \right)\left( {1 - {P_{{{\mathbb{A}}_{\rm{n}}}}}} \right)} \right).
	\end{align}
	%\frac{2^{\eta _b-1}}{2^{\eta _b}-1} where 0.5 means that only half of the bits are faulty for each symbol error.
	The final BER of the proposed MCR-IM system under Nakagami-${m}$ fading channels can be computed as
	\begin{align}
		\label{final nakagami}
		{P_{{\rm{err}}}} = \int_0^\infty  {{P_\alpha }{f_m}\left( \alpha  \right)d\alpha }  = \int_{ - \infty }^\infty  {{P_{{e^x}}}{f_m}\left( {{e^x}} \right){e^x}dx}.
	\end{align}

	We evaluate (\ref{final nakagami}) using the Gauss-Hermite quadrature approach. According to Table. (25.10) in \cite{abramowitz1948handbook}, one has
	\begin{align}
		\label{eq201}
		\int_{ - \infty }^\infty  {g\left( x \right)dx}  = \sum\limits_{w = 1}^W {{\varphi _w}g\left( {{x_w}} \right)\exp \left( {x_w^2} \right)}  + {O_W},
	\end{align}
	where ${W}$ denotes the number of sample points used for the approximation, ${{{x_w}}}$ is the ${w}$-th root of the Hermite polynomial ${{H_W}\left( x \right)}$, ${w = 1,2, \cdots ,W}$, ${{\varphi _w}}$ is the ${w}$-th associated weight obtained from ${\frac{{{2^{W - 1}}W!\sqrt \pi  }}{{{W^2}H_{W - 1}^2\left( {{x_w}} \right)}}}$ and ${{O_W}}$ is residual term, which tends to ${0}$ when ${W}$ tends to infinity.
	Then, (\ref{final nakagami}) can be further calculated as
	\begin{align}
		{P_{{\rm{err}}}} = \sum\limits_{w = 1}^W {{\varphi _w}{P_{{e^{{x_w}}}}}{f_m}\left( {{e^{{x_w}}}} \right)\exp \left( {{x_w} + x_w^2} \right)}  + {O_W}.
	\end{align}
	
	It is noted that the precise BER representation, involving confluent hypergeometric functions ${{}_1{F_1}\left( {a,b,x} \right)}$, is computationally expensive.
	To reduce computational complexity while maintaining accuracy, we approximate the mean and variance in the formula to reduce the amount of computation and complexity.
	For the signal set ${{{\mathbb{A}}_{\rm{s}}}}$ and interference set ${{{\mathbb{A}}_{\rm{i}}}}$, (\ref{mu}) and (\ref{sigma}) can be approximated by ${{\mu _{\ell ,k}} \approx \left| {h{{\bf{u}}_\ell }\left( k \right)} \right|}$ and ${\sigma _{\ell ,k}^2 \approx \frac{{{N_0}}}{2}}$, respectively.
	For the noise set ${{{\mathbb{A}}_{\rm{n}}}}$, since ${{{\bar U}_n}}$ is relatively small, it becomes negligible after squaring, leading to ${\bar U_n^2 \approx 0}$.
	Finally, the closed-form BER expression of the proposed MCR-IM system is derived in (\ref{new_final BER}), on the top of the next page.
	\begin{figure*}[ht]
		\begin{align}
			\label{new_final BER}
			{P_{{\rm{err}}}} \approx \frac{{{m^m}}}{{\Gamma \left( m \right)}}\sum\limits_{w = 1}^W {\left( {{\varphi _w}\left( \begin{array}{l}
						1 - \left( {\prod\limits_{a = 1}^{\left| {{{\mathbb{A}}_{\rm{s}}}} \right|} {\prod\limits_{b = 1}^{\left| {{{\mathbb{A}}_{\rm{i}}}} \right|} {Q\left( {\frac{{{\mu _{{{\mathbb{A}}_{\rm{i}}}\left( b \right)}} - {\mu _{{{\mathbb{A}}_{\rm{s}}}\left( a \right)}}}}{{\sqrt {{N_0}} }}} \right)} } } \right) \times \\
						\left( {\prod\limits_{a = 1}^{\left| {{{\mathbb{A}}_{\rm{s}}}} \right|} {Q\left( {\sqrt {2{H_{\left| {{{\mathbb{A}}_{\rm{n}}}} \right|}}}  - \sqrt {\frac{2}{{{N_0}}}} {\mu _{{{\mathbb{A}}_{\rm{s}}}\left( a \right)}}} \right)} } \right)
					\end{array} \right){e^{\left( {2m{x_w} - m{e^{2{x_w}}} + x_w^2} \right)}}} \right)}  + {O_W}.
		\end{align}
		{\noindent} \rule[-10pt]{18cm}{0.05em}
	\end{figure*}
	
	\section{Multi-user Detection}
	In this section, we propose a PD-SIC algorithm for multi-user detection, where this algorithm is performed at the gateway.
	
	\subsection{Preamble Detection}
	
	\begin{table}[t]
		\centering							% 居中
		\renewcommand{\arraystretch}{2}		% 调整表格高度
		\setlength{\tabcolsep}{7pt}			% 调整表格宽度
		\caption{Range of chirp rates for different users and the adopted chirp rates for the preamble.}
		\label{table:Chirp Rate Range}
		\begin{tabular}{c|c|c}		% 标注多少列，哪些列不要竖线
			\Xhline{1pt}
			User & Chirp Rate Range (${{M_1}\to{M_P}}$) & Preamble Chirp Rate\\
			\hline
			User 1 & ${1 \to P}$ & 1\\
			\hline
			User 2 & ${P + 1 \to 2P}$ & ${P + 1}$\\
			\hline
			${ \cdots }$ & ${ \cdots }$ & ${ \cdots }$\\
			\hline
			User ${{N_u}}$ & ${\left( {{N_u}  - 1} \right)P + 1 \to {N_u}P}$ & ${\left({{N_u}- 1} \right)P + 1}$\\
			\Xhline{1pt}
		\end{tabular}
	\end{table}
	
	\begin{figure*}[t]
		\centering
		\includegraphics[scale=0.7]{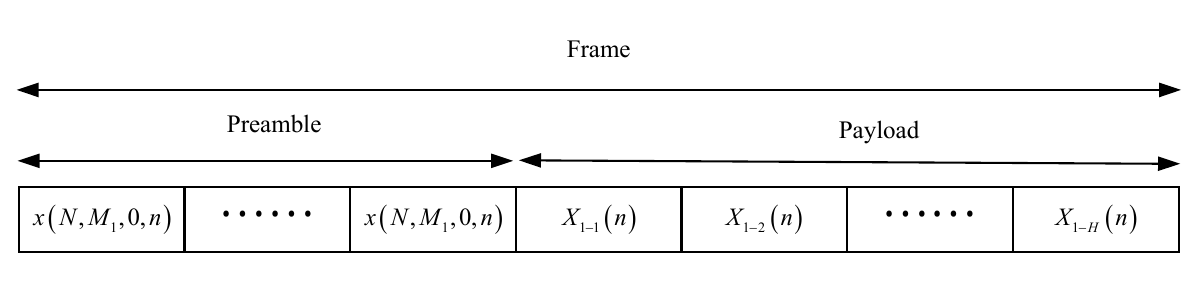}
		\caption{Frame structure of the proposed MCR-IM system.}
		\label{fig16}
	\end{figure*}
	
	In traditional LoRa systems, a data packet consists of a preamble and a payload.
	A LoRa frame contains a variable-length preambles, a sync word, and a start frame delimiter (SFD), which facilitate packet detection and symbol synchronization.
	In the MCR-IM system, we adopt a frame structure similar to that of LoRa.
	However, different users adopt a selectable minimum chirp rate as the preamble.
	We consider ${{N_u}}$ users transmitting signals simultaneously.
	The chirp rate selection range for different users is shown in Table \ref{table:Chirp Rate Range}.
	User 1 uses up-chirp with chirp rate 1 as the preamble selection, while user 2 uses chirp rate ${P+1}$.
	At the receiver, only the corresponding minimum chirp rate needs to be demodulated to perform packet detection and symbol synchronization.
	
	The frame format of the proposed MCR-IM system is shown in Fig.~\ref{fig16}.
	By assigning different preamble chirp rates to different users, the receiver can independently detect different users during packet detection and symbol synchronization \cite{10360184}.
	This minimizes inter-user interference, enabling the simultaneous detection of multiple data packets.
	In addition, through the detection of preamble signals with different chirp rates, the gateway can obtain the number of concurrent users and the corresponding chirp rate of the users.
	This will provide data detection for the proposed PD-SIC algorithm.
	
	\subsection{PD-SIC Algorithm}
	
	\begin{figure}[t]
		\centering
		\includegraphics[scale=0.6]{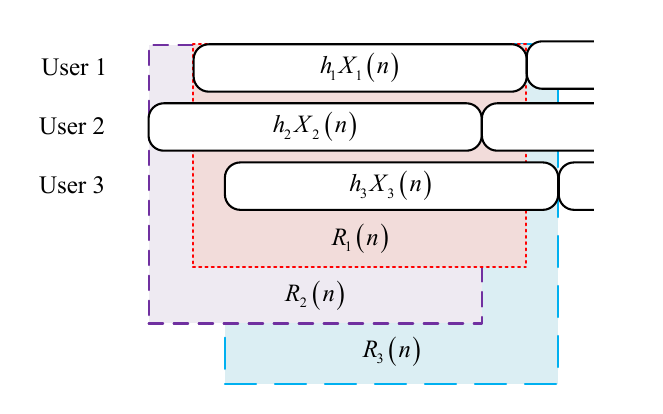}
		\caption{The received signal is divided into different segments when ${{N_u} = 3}$.}
		\label{multi_user_03}
	\end{figure}
	
	In traditional LoRa systems, when two signals collide, the receiver typically captures the signal with the higher peak energy.
	When peak energies among different users are similar, the receiver struggles to differentiate between them, which  inevitably leads to data loss.
	In contrast, the MCR-IM system assigns different chirp rate ranges to different users, allowing detection of signals with their corresponding chirp rates.
	The basic idea behind the PD-SIC algorithm is to determine the phase of the cancellation signal by detecting the peak energy across different phase cases.
	Then, users' signals are sequentially estimated and eliminated.
	
	We consider a total of ${{N_u}}$ users transmitting simultaneously.
	Symbol collisions among different users are not perfectly aligned over time.
	Based on the preamble detection process described in the previous subsection, we assume that the arrival times of signals from different users are known at the receiver.
	Fig.~\ref{multi_user_03} illustrates that the received signal is divided into different segments according to the arrival time of the user.
	Furthermore, the channel fading experienced by each user follows an independent Nakagami-${m}$ distribution.

	The PD-SIC algorithm is given in detail as follows:
	\begin{itemize}
		\item [1)]
		Using (\ref{demo_01}) and (\ref{demo_02}), we directly demodulate the received signal to obtain the corresponding ${{\bf{Y}}}$ matrix for each user, i.e.,${\left\{ {{{\bf{Y}}_1},{{\bf{Y}}_2}, \cdots ,{{\bf{Y}}_{N_u} }} \right\}}$.
		The sum of the maximum ${L}$ peak energies for each user is estimated by
		\begin{align}
			{E_i} = \sum\limits_{l = 1}^L {{{\max }^{\left( l \right)}}\left( {{{\left| {{{\bf{Y}}_i}} \right|}^2}} \right)}.
		\end{align}
		where ${{\max ^{\left( l \right)}}\left(  \cdot  \right)}$ returns the ${l}$-th largest element in the matrix ${{\left| {{{\bf{Y}}_i}} \right|^2}}$.
		The user with the highest total peak energy is given priority for demodulation.
		\item [2)]
		We assume that ${\chi  = \mathop {\max }\limits_i \left\{ {{E_i}|i = 1,2, \cdots ,{N_u} } \right\}}$.
		From (\ref{demo 23})-(\ref{demo 25}), we can obtain the indexes information ${\left( {{{{\bf{\hat l}}}_\chi },{{{\bf{\hat k}}}_\chi }} \right)}$ of the ${L}$ peak points in ${{{\bf{Y}}_\chi }}$, and the estimated transmitted symbol ${{{{\bf{\tilde s}}}_\chi }}$ of user ${{\chi }}$.
		Then, we construct the signal ${{X_\chi }\left( n \right)}$ based on the estimated symbol ${{{{\bf{\tilde s}}}_\chi }}$ and energy ${{E_\chi }}$, as defined in (\ref{modu 8})-(\ref{modu 11}).
		
		\item [3)]
		The phase is divided into ${K}$ segments and one has
		\begin{align}
			{\theta _d} = \frac{{2\pi }}{K}d,
		\end{align}
		where ${d \in \left\{ {0,1, \cdots ,K - 1} \right\}}$.
		We first test the case of ${d=0}$. The signal after cancellation is given by
		\begin{align}
			\label{Td}
			{T_d}\left( n \right) = {R_\chi }\left( n \right) - \exp \left( {j{\theta _d}} \right){X_\chi }\left( n \right),
		\end{align}
		where ${{R_\chi }\left( n \right)}$ represents the portion of the signal corresponding to user ${\chi }$ when ${{N_u}}$ users collide.
		We then recalculate the ${{{{\bf{\hat Y}}}_\chi }}$ corresponding to ${{T_d}\left( n \right)}$, given by
		\begin{align}
			\label{Rx}
			{{{\bf{\hat Y}}}_\chi } = {\left[ {{{{\bf{\hat y}}}_{\left( {\chi  - 1} \right)P + 1}};{{{\bf{\hat y}}}_{\left( {\chi  - 1} \right)P + 2}}; \cdots ;{{{\bf{\hat y}}}_{\chi P}}} \right]^{\rm{T}}},
		\end{align}
		where
		\begin{align}
			\label{Rl}
			{{{\bf{\hat y}}}_\ell }\left( k \right) = {\rm{DFT}}\left( {{T_d}\left( n \right)\frac{1}{{\sqrt N }}{e^{\left( { - j\frac{\pi }{N}\ell \left( {n + 1} \right)n} \right)}}} \right).
		\end{align}
	
		Since we have obtained ${\left( {{{{\bf{\hat l}}}_\chi },{{{\bf{\hat k}}}_\chi }} \right)}$ in step 2), we calculate the sum of original peak energy after canceling the signal ${\exp \left( {j{\theta _d}} \right){X_\chi }\left( n \right)}$ as follows
		\begin{align}
			{{{\hat E}_\chi } = \sum {{{\left| {{{{\bf{\hat Y}}}_\chi }\left( {{{{\bf{\hat l}}}_\chi },{{{\bf{\hat k}}}_\chi }} \right)} \right|}^2}} }.
		\end{align}
		
		\item [4)]
		Based on ${K}$, we can obtain the cancellation strength coefficient ${\beta }$, given by
		\begin{align}
			\label{beta}
			\beta  = {\left( {2\sin \frac{\pi }{{2K}}} \right)^2} = 2\left( {1 - \cos \frac{\pi }{K}} \right).
		\end{align}
		The condition ${{{\hat E}_\chi } < \beta {E_\chi }}$ is used to determine if the phase selection is correct.
		If it is not satisfied, then we move to the next phase.
		After canceling the interference from users with stronger peak energy, we set ${{E_\chi } = 0}$, and return to step 2) to continue the demodulation of users' signals with lower energy, until the demodulation is completed for all the users.
	\end{itemize}
	
	Since symbols are often not aligned across different users, we assume that the sum of peak energies obtained in step 1) satisfies ${{E_1} > {E_2} >  \cdots  > {E_{N_u} }}$.
	Then, the demodulation order is user 1${\to}$user 2${\to}$${\cdots}$${\to}$user ${{N_u}}$.
	The demodulation of users with higher peak energy must be completed before that of users with lower peak energy.
	The specific detailed steps of the proposed PD-SIC are shown in Algorithm \ref{alg:algorithm01}.
	As ${K}$ increases, the PD-SIC algorithm performs better and requires more computation.
	It should be noted that the condition ${{{\hat E}_\chi } < \beta {E_\chi }}$ may not be satisfied when the SNR is very low.
	This is because noise may cause different degrees of phase shifts at different peak points.
	
	\begin{algorithm}[t]
		\SetAlgoLined 									% 显示行号
		\caption{PD-SIC algorithm (${{N_u} > 1}$)} 		% 算法标题
		\label{alg:algorithm01}
		\KwIn{${\nu}$, ${N}$, ${P}$, ${L}$, ${{N_u}}$, ${{R_1}\left( n \right),{R_2}\left( n \right), \cdots ,{R_{{N_u}}}\left( n \right)}$} 	% 输入参数
		\KwOut{${{{{\bf{\tilde s}}}_1};{{{\bf{\tilde s}}}_2}; \cdots ;{{{\bf{\tilde s}}}_{N_u} }}$} 	% 输出参数
		${{I_{dx}} = \mathbf{0}}$; ${K=8}$; ${{E_{\max }} = 0}$; \\
		\For{${i = 1:1:{N_u} }$}
		{
			\For{${\ell  = \left( {i - 1} \right)P + 1:1:iP}$}
			{
				${{{\bf{y}}_\ell }\left( k \right) = {\rm{DFT}}\left( {{R_i}\left( n \right)\frac{1}{{\sqrt N }}{e^{\left( { - j\frac{\pi }{N}\ell \left( {n + 1} \right)n} \right)}}} \right)}$;
			}
			${{{\bf{Y}}_i} = {\left[ {{{\bf{y}}_{\left( {i - 1} \right)P + 1}};{{\bf{y}}_{\left( {i - 1} \right)P + 2}}; \cdots ;{{\bf{y}}_{iP}}} \right]^{\rm{T}}}}$;\\
			${{E_i} = \sum\limits_{l = 1}^L {{{\max }^{\left( l \right)}}\left( {{{\left| {{{\bf{Y}}_i}} \right|}^2}} \right)} }$;\\
		}
		\For{${u = 1:1:{N_u} }$}
		{
			${\chi  = \mathop {\max }\limits_i \left\{ {{E_i}|i = 1,2, \cdots ,{N_u} } \right\}}$;\\
			\If{${u>1}$}
			{
				\For{${\ell  = \left( {\chi  - 1} \right)P + 1:1:\chi P}$}
				{
					${{{\bf{y}}_\ell }\left( k \right) = {\rm{DFT}}\left( {{R_u}\left( n \right)\frac{1}{{\sqrt N }}{e^{\left( { - j\frac{\pi }{N}\ell \left( {n + 1} \right)n} \right)}}} \right)}$;
				}
				${{{\bf{Y}}_\chi } = {\left[ {{{\bf{y}}_{\left( {\chi  - 1} \right)P + 1}};{{\bf{y}}_{\left( {\chi  - 1} \right)P + 2}}; \cdots ;{{\bf{y}}_{\chi P}}} \right]^{\rm{T}}}}$;\\
			}
			
			${\left( {{{{\bf{\hat l}}}_\chi },{{{\bf{\hat k}}}_\chi }} \right) = \arg \mathop {\max }\limits_{{\bf{l}},{\bf{k}}} \left| {{{\bf{Y}}_\chi }} \right|}$;\\
			${{{{\bf{\tilde s}}}_\chi } = {\rm{sort}}\left( {\left( {{{{\bf{\hat l}}}_\chi } - 1} \right){2^\nu } + {{{\bf{\hat k}}}_\chi } - 1} \right)}$;\\
			\If{${u =  = {N_u} }$}
			{break;\\}
			Generate ${{X_\chi }\left( n \right)}$ with signal energy ${{E_\chi }}$;\\
			\For{${d = 0:1:K - 1}$}
			{
				${{\theta _d} = \frac{{2\pi }}{K}d}$;\\
				${{T_d}\left( n \right) = {R_\chi }\left( n \right) - \exp \left( {j{\theta _d}} \right){X_\chi }\left( n \right)}$;\\
				\For{${\ell  = \left( {\chi  - 1} \right)P + 1:1:\chi P}$}
				{
					${{{{\bf{\hat y}}}_\ell }\left( k \right) = {\rm{DFT}}\left( {{T_d}\left( n \right)\frac{1}{{\sqrt N }}{e^{\left( { - j\frac{\pi }{N}\ell \left( {n + 1} \right)n} \right)}}} \right)}$;
				}
				${{{{\bf{\hat Y}}}_\chi } = {\left[ {{{{\bf{\hat y}}}_{\left( {\chi  - 1} \right)P + 1}};{{{\bf{\hat y}}}_{\left( {\chi  - 1} \right)P + 2}}; \cdots ;{{{\bf{\hat y}}}_{\chi P}}} \right]^{\rm{T}}}}$;\\
				${{{\hat E}_\chi } = \sum {{{\left| {{{{\bf{\hat Y}}}_\chi }\left( {{{{\bf{\hat l}}}_\chi },{{{\bf{\hat k}}}_\chi }} \right)} \right|}^2}} }$;\\
				\If{${{{\hat E}_\chi } < \beta {E_\chi }}$}
				{
					${{R_\chi }\left( n \right) = {T_d}\left( n \right)}$;\\
					break;\\
				}
			}
			${{E_\chi } = 0}$;\\
		}
	\end{algorithm}

	\subsection{Algorithmic Complexity}
	
	\begin{table*}[t]
		\centering							% 居中
		\renewcommand{\arraystretch}{2}		% 调整表格高度
		\setlength{\tabcolsep}{7pt}			% 调整表格宽度
		\caption{Complexity of direct demodulation algorithm and PD-SIC algorithm for the proposed MCR-IM system.}
		\label{table:complexity}
		\begin{tabular}{|c|c|c|c|}		% 标注多少列，哪些列不要竖线
			\Xhline{1pt}
			& Direct Demodulation Algorithm
			& PD-SIC Algorithm (${{N_u}  > 1}$) \\
			\hline
			Dechirp
			& ${P}$
			& ${\frac{P}{2}\left( {K + 5 - \frac{{K + 3}}{{N_u}}} \right)}$ \\
			\hline
			DFT
			& ${P}$
			& ${\frac{P}{2}\left( {K + 5 - \frac{{K + 3}}{{N_u}}} \right)}$ \\
			\hline
			Metrix max
			& ${1}$
			& ${2}$ \\
			\hline
			MUL
			& ${2PN\left( {2 + {{\log }_2}N} \right)}$
			& ${P\left( {K + 5 - \frac{{K + 3}}{{N_u}}} \right)N\left( {2 + {{\log }_2}N} \right)}$ \\
			\hline
			ADD	
			& ${PN\left( {2 + 3{{\log }_2}N} \right)}$
			& ${\frac{P}{2}\left( {K + 5 - \frac{{K + 3}}{{N_u}}} \right)N\left( {2 + 3{{\log }_2}N} \right)}$ \\
			\hline
			CMP
			& ${P2^{\nu} - L + \left( {L - 1} \right){\log_2}\left( {P2^{\nu}} \right)}$
			& ${2\left( {P{2^{\nu}} - L + \left( {L - 1} \right){{\log }_2}\left( {P{2^{\nu}}} \right)} \right)}$\\
			\Xhline{1pt}
		\end{tabular}
	\end{table*}
	
	The computational complexity of the PD-SIC algorithm is proportional to ${K}$.
	For ${{N_u}}$ colliding users, the receiver must performs ${P{N_u}}$ dechirp and DFT operations on the received signal and ${{N_u}}$ peak-search operations.
	The receiver then selects the signal with the largest peak energy for demodulation.
	From the demodulation results and peak energies, the cancellation signal ${{X_\chi }\left( n \right)}$ can be constructed.
	On average of ${\frac{{K + 1}}{2}}$ phase attempts are required to determine the phase of ${{X_\chi }\left( n \right)}$.
	Each phase attempt requires ${P}$ dechirp and DFT operations.
	For ${{N_u}}$ users, we need to perform ${{N_u}-1}$ signal cancellations.
	After each signal cancellation, we demodulate the next signal, which also requires ${P}$ dechirp and DFT operations and one peak-search operation.
	
	In summary, for ${{N_u}}$ users, the total number of dechirp and DFT operations is ${\frac{P}{2}\left( {\left( {K + 5} \right){N_u} - K - 3} \right)}$.
	The total number of peak-search operations is ${2{N_u}-1}$.
	Each user requires ${\frac{P}{2}\left( {K + 5 - \frac{{K + 3}}{{N_u}}} \right)}$ dechirp and DFT operations, and 2 peak-search operations.
	
	To allow comparison with other schemes, we have converted the above results into the number of multipliers (MUL), adders (ADD) and comparators (CMP).
	Table \ref{table:complexity} presents the number of operations for the direct demodulation and PD-SIC algorithms.
	In addition, complexity comparisons for more systems can be found in \cite{9542955}.
	Since the algorithm is performed at the gateway, the computational load per symbol being proportional to ${P}$ and ${K}$ is acceptable in practical scenarios.

	\section{Results and Discussions}
	
	In this section, the performance of the proposed  MCR-IM system is evaluated with both direct demodulation and PD-SIC algorithms.
	For performance comparison, we select the FSCSS-IM system, which also uses the indexing technique, and the GCSS system, which uses the group indexing technique.
	For multi-user detection, we compare against the OrthoRa system, which also exploits orthogonality to achieve parallel transmission.
	
	\subsection{Performance Evaluation of the MCR-IM System}
	
	\begin{figure}[t]
		\centering
		\includegraphics[scale=0.32]{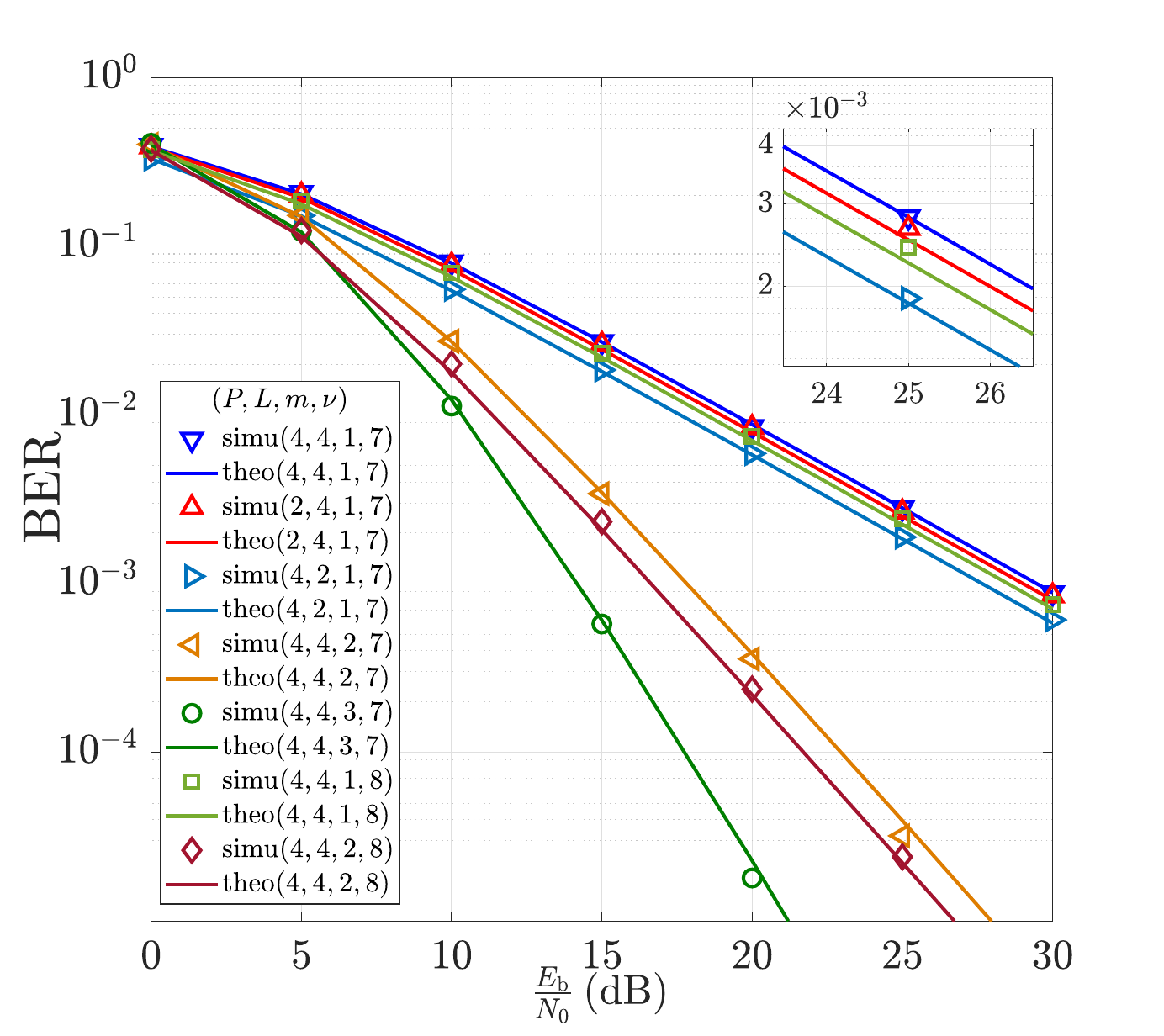}
		\caption{Theoretical and simulated BER of the proposed MCR-IM system with the direct demodulation algorithm over Nakagami-${m}$ fading channels.}
		\label{fig03}
	\end{figure}
	\begin{figure}[t]
		\centering
		\includegraphics[scale=0.32]{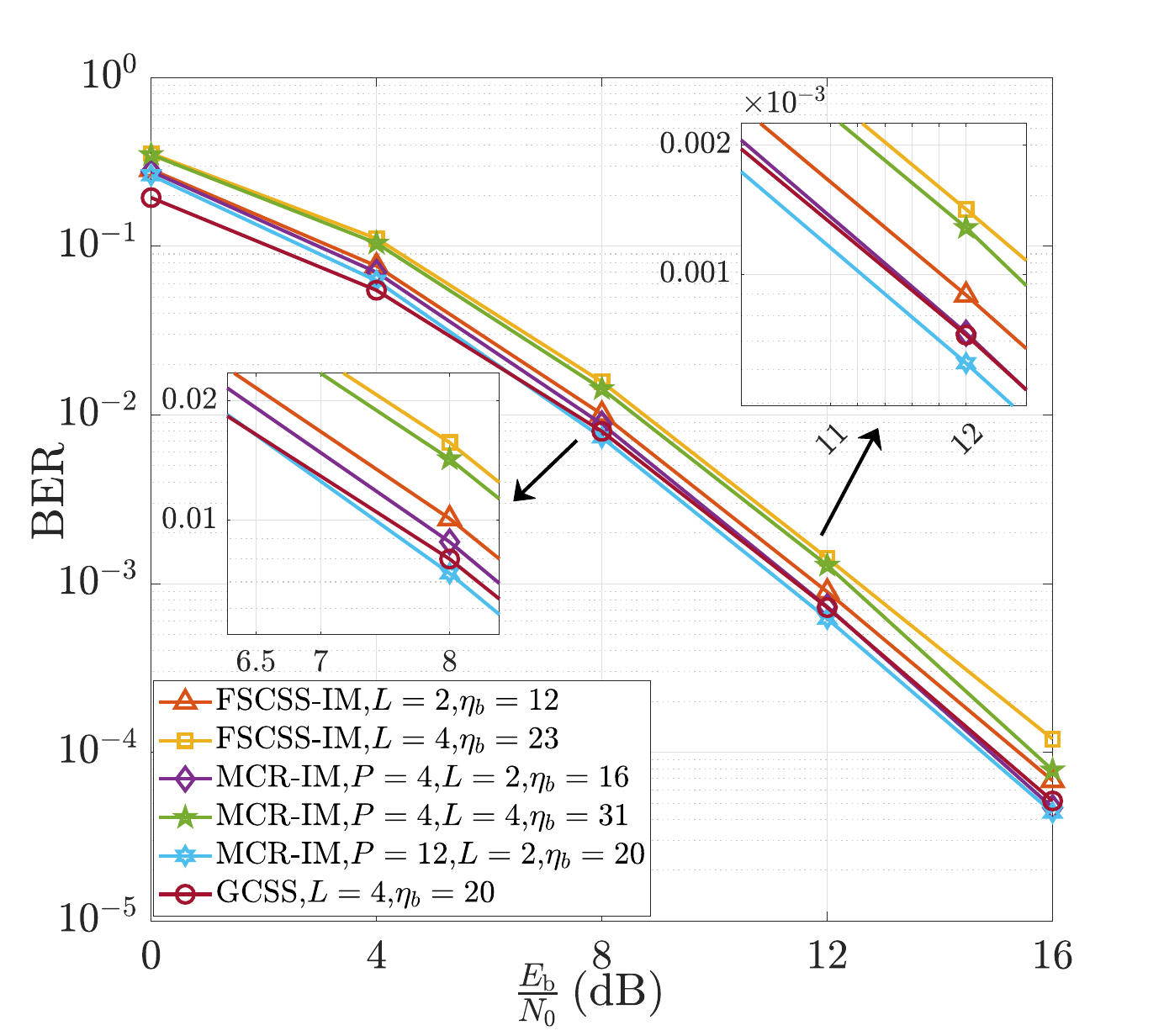} % AWGN is fig12.pdf
		\caption{BER performance of the MCR-IM, FSCSS-IM, and GCSS systems over Nakagami-${m}$ fading channels, with ${\nu=10}$ and ${m=3}$.}
		\label{fig12}
	\end{figure}
	\begin{figure}[t]
		\centering
		\includegraphics[scale=0.32]{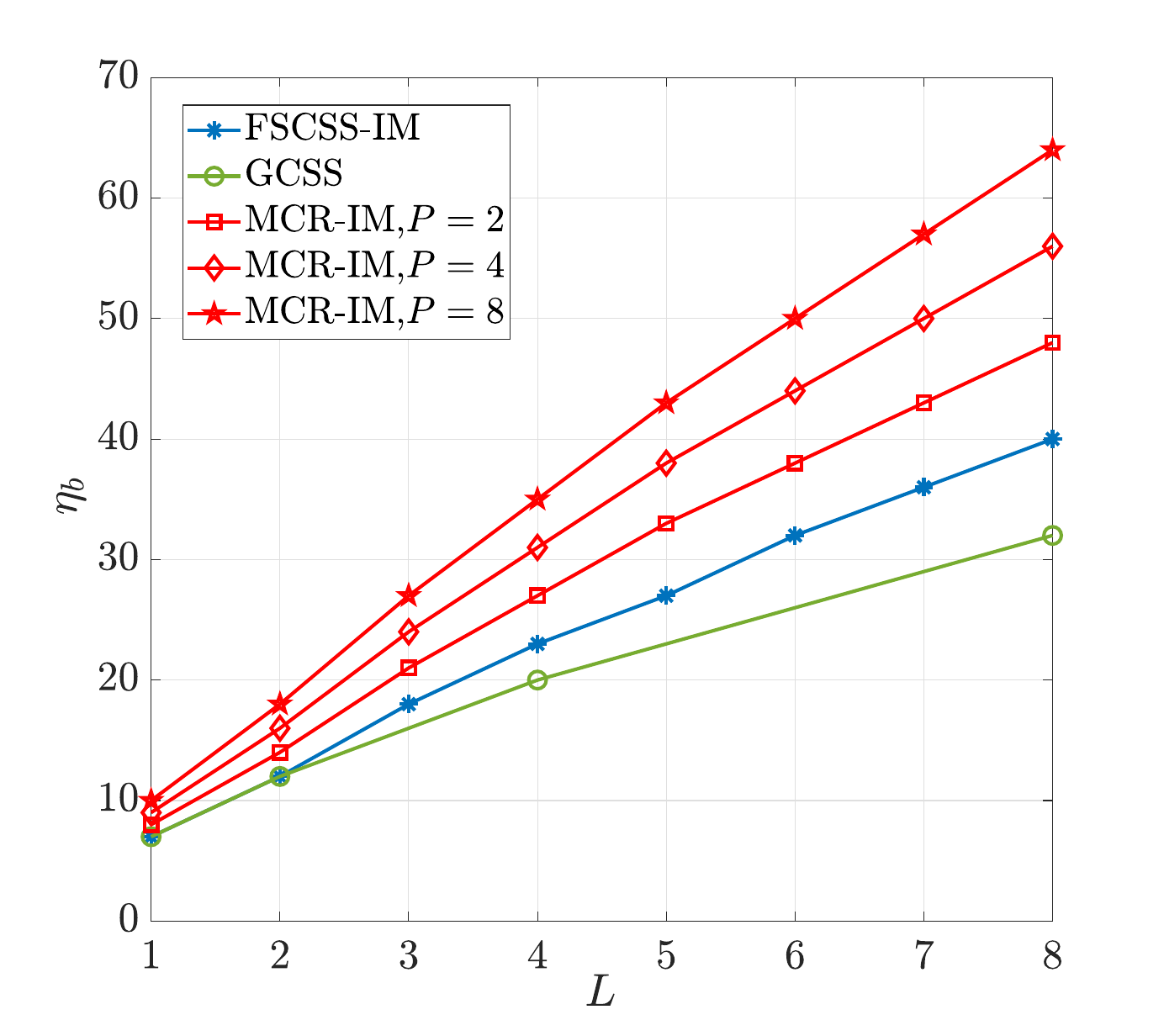}
		\caption{Number of transmitted bits per symbol ${{\eta _b}}$ for the MCR-IM, FSCSS-IM, and GCSS systems considering different numbers of indexes ${L}$, with ${\nu=7}$.}
		\label{fig11}
	\end{figure}
	\begin{figure}[t]
		\centering
		\includegraphics[scale=0.32]{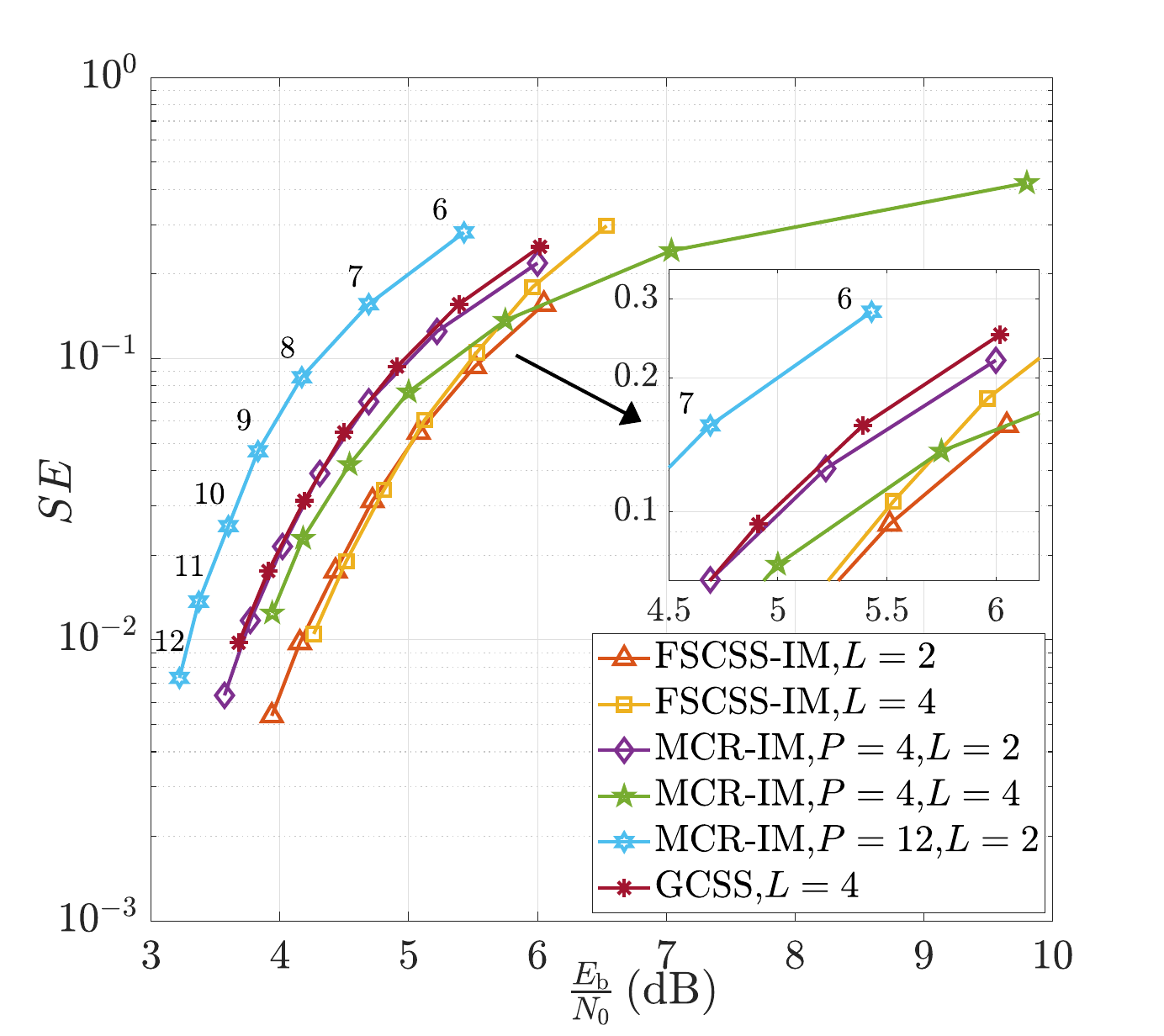}
		\caption{SE of the GCSS, FSCSS-IM, and MCRIM systems, over an AWGN channel with non-coherent demodulation, and ${\nu = 6,7, \cdots ,12}$.}
		\label{fig04}
	\end{figure}
	
	Fig.~\ref{fig03} presents the theoretical and simulated BER performance of the MCR-IM system with the direct demodulation algorithm.
	The results show that the theoretical derivations perfectly match the simulated outcomes, thereby validating the derived analytical expressions.
	With the shape parameter ${m=1}$ (i.e., under the Rayleigh fading channel), ${\nu=7}$ and ${L=4}$, the SNR performance loss is less than 0.5 dB when ${P}$ is increased from 2 to 4.
	This increases the number of bits carried per symbol from 27 to 31.
	Reducing ${L}$ from 4 to 2 when ${m=1}$, ${P=4}$, and ${\nu=7}$, yields a gain of about 2 dB in the BER performance of the proposed system.
	The trade-off is a reduction in the number of bits per symbol from 31 to 16.
	When ${P=L=4}$, ${m=1}$, and ${\nu}$ is increased from 7 to 8, an improvement in the BER performance of approximately ${1}$ dB is obtained.
	
	Fig.~\ref{fig12} compares the BER performance of the MCR-IM, FSCSS-IM, and GCSS systems over Nakagami-$m$ fading channels.
	It should be noted that ${{\eta _b}}$ is before defined as the number of bits carried per symbol.
	When ${L=2}$, the MCR-IM system with ${P=4}$ and ${P=12}$ shows an SNR gain of 0.4 dB and 0.6 dB compared to the FSCSS-IM system, respectively, while also carrying more bits per symbol than the FSCSS-IM system.
	This advantage can be attributed to the utilization of multiple chirp rates.
	When ${L=4}$, the MCR-IM system with ${P=4}$ shows an SNR gain of 0.2 dB compared to the FSCSS-IM system.
	The MCR-IM system with ${L=2}$ and ${P=12}$ has the same ${{\eta _b}}$ as the GCSS system with ${L=4}$.
	It is noted that the BER performance of the MCR-IM system surpasses that of GCSS system when ${\frac{{{E_{\rm{b}}}}}{{{N_0}}} > 6.5}$ dB, with a gain reaching 0.2 dB.
	
	Fig.~\ref{fig11} compares the number of transmitted bits per symbol ${{\eta _b}}$ of the MCR-IM, FSCSS-IM, and GCSS systems considering different number of indexes ${L}$, where ${\nu=7}$.
	It is noted that the MCR-IM system can transmit more information bits with an equivalent number of chirps compared to other multiple chirp systems.
	For ${L=4}$ and ${P=4}$, the number of bits carried by the MCR-IM systems is ${55\% }$ and ${34\%}$ more than that of the GCSS and FSCSS-IM system, respectively.
	It is worth noting that FSCSS-IM can be considered as an MCR-IM system with ${P=1}$.
	When ${L=8}$, as ${P}$ increases to 2, 4, and 8, the number of bits carried by the MCR-IM system increases by ${20\%}$, ${40\%}$, and ${60\%}$ compared to ${P=1}$, respectively.
	The MCR-IM system has flexible transmission rate options, making it well suited for various IoT applications.

	Fig.~\ref{fig04} shows the SE of the GCSS, FSCSS-IM, and MCR-IM systems, over an AWGN channel and considering non-coherent demodulation.
	For the SE analysis, we assume a bandwidth of ${B = 125{\rm{kHz}}}$ and a sampling interval of ${{T_{\rm{c}}} = \frac{1}{B}}$.
	Thus, the duration of a symbol is ${{T_{\rm{s}}} = {2^{\nu}}{T_{\rm{c}}}}$.
	The SE of the different systems can be obtained as ${SE = \frac{{{\eta _b}\left( {1 - {P_{{\rm{err}}}}} \right)}}{{{T_{\rm{s}}}B}}}$.
	It is observed that, compared to the FSCSS-IM system with ${L = 2}$, the MCR-IM system with ${P=4}$ and ${L=2}$ requires lower SNR value to reach a targeting SE.
	Moreover, for a fixed SNR, the MCR-IM system achieves a higher SE.
	When ${L = 2}$, the curves from Fig.~\ref{fig04} undergo a notable upward-left shift as ${P}$ increases from 1 to 4, and further to 12.	
	It is shown that when the SF is small, the increase in ${P}$ primarily results in an improvement of the SE, thereby accelerating the transmission rate.
	Conversely, for a large SF, while the increment in ${P}$ yields only a marginal improvement of the SE, it enables the achievement of a lower SNR.
	Moreover, one can observe that increasing ${L}$ can improve the SE, but also increase SNR requirements, especially when the SF is small.
	
	\subsection{Performance Evaluation of Multi-User Detection }
	
	\begin{figure}[t]
		\centering
		\includegraphics[scale=0.32]{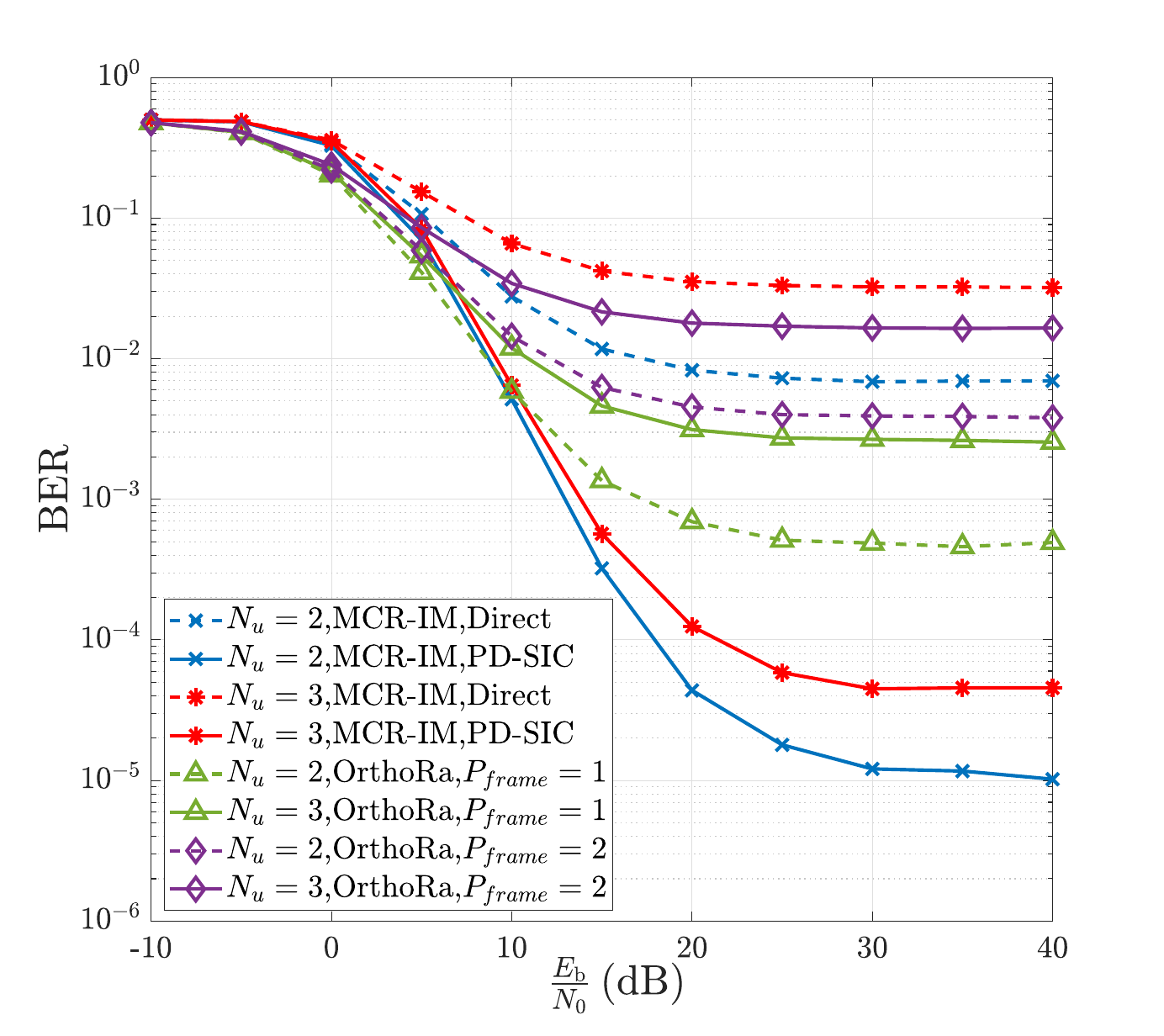}
		\caption{BER performance of the MCR-IM and OrthoRa systems over Nakagami-$m$ fading channels, for ${\nu=7}$, ${P=4}$, ${L=2}$, ${K=8}$ and ${m=3}$.}
		\label{fig15}
	\end{figure}
	
	\begin{figure}[t]
		\centering
		\includegraphics[scale=0.32]{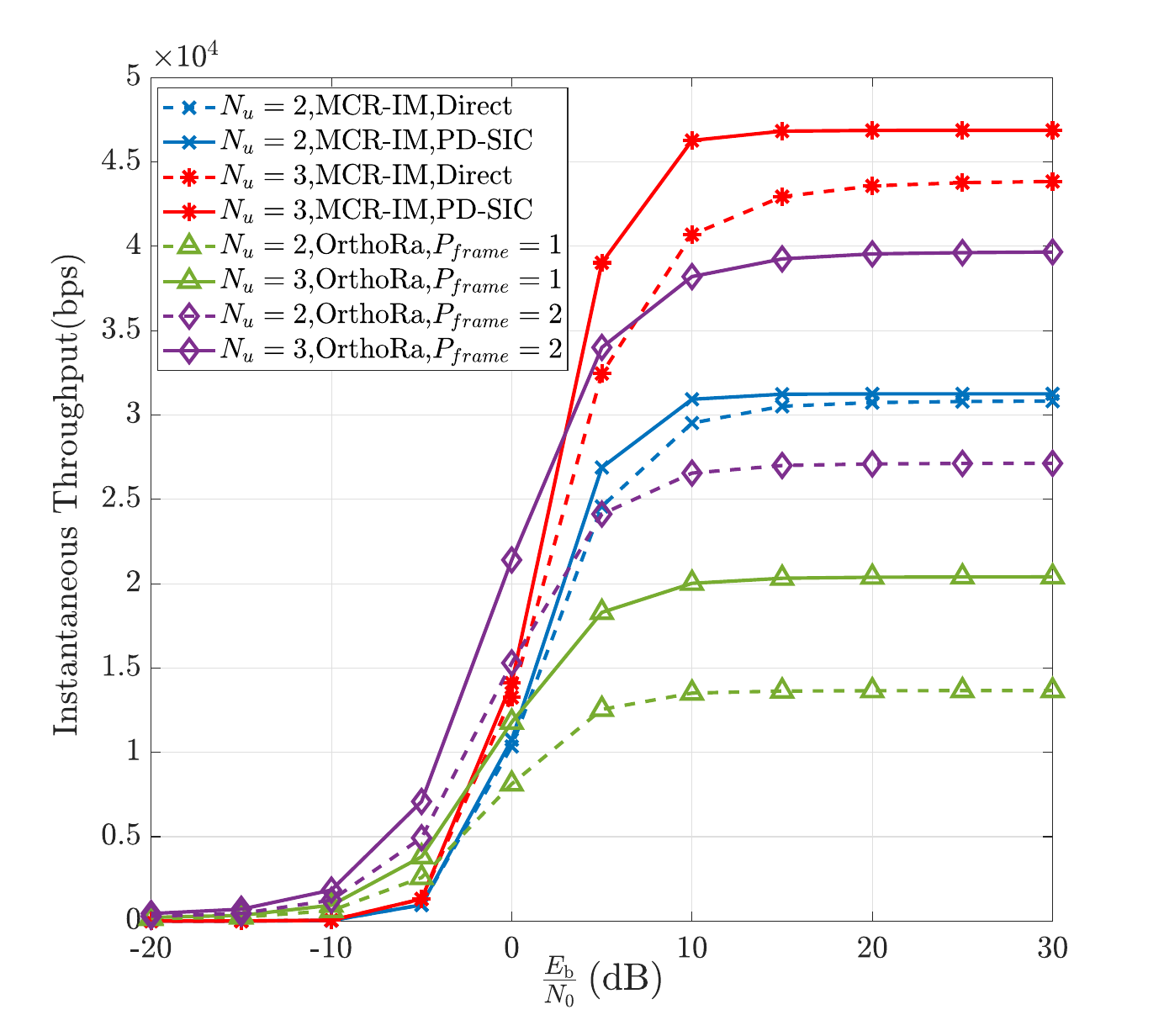}
		\caption{Instantaneous throughput of the MCR-IM and OrthoRa systems over Nakagami-$m$ fading channels, for ${\nu=7}$, ${P=4}$, ${L=2}$, ${K=8}$ and ${m=3}$.}
		\label{fig14}
	\end{figure}
	
	Fig.~\ref{fig15} shows the BER of MCR-IM and OrthoRa systems. Here, the MCR-IM system adopts both direct demodulation and PD-SIC algorithms.
	The OrthoRa system adopts non-coherent demodulation, where ${{P_{frame}}}$ represents the number of parallel frames.
	Similar to the MCR-IM system, the OrthoRa system achieves quasi-orthogonality by utilizing scattering chirps.
	Extensive literature research indicates that the number of colliding users in LoRaWAN is generally 2, and rarely 3\cite{9131834}.
	As shown in Fig.~\ref{fig15}, compared to the direct demodulation algorithm, the PD-SIC algorithm enables the MCR-IM system to achieve a lower BER in case of collisions.
	When ${\frac{{{E_{\rm{b}}}}}{{{N_0}}} < 5}$ dB, the OrthoRa system exhibits slightly better BER performance than the MCR-IM system.
	However, when ${\frac{{{E_{\rm{b}}}}}{{{N_0}}} > 5}$ dB, the MCR-IM system with the PD-SIC algorithm outperforms the OrthoRa system significantly, while the OrthoRa system outperforms the MCR-IM system with the direct demodulation algorithm.
	
	Fig.~\ref{fig14} depicts the instantaneous throughput of the MCR-IM and OrthoRa systems.
	Similarly, we assume a bandwidth of ${B = 125{\rm{kHz}}}$ and a sampling interval of ${{T_{\rm{c}}} = \frac{1}{B}}$. Thus, the duration of a symbol is ${{T_{\rm{s}}} = {2^{\nu}}{T_{\rm{c}}}}$.
The instantaneous throughput is given by ${{T_{throughput}} = \frac{{{\eta _b}}}{{{T_{\rm{s}}}}}\left( {1 - {P_{err\left| {{N_u}} \right.}}} \right){N_u}}$, where ${{{P_{err\left| {{N_u}} \right.}}}}$ represents the BER under the collision of ${{N_u}}$ users.
	The results indicate that the MCR-IM system with the direct demodulation algorithm achieves a high instantaneous throughput during collisions.
	Furthermore, applying the PD-SIC algorithm enhances the instantaneous throughput of the MCR-IM system, particularly when the number of colliding users (i.e., ${{N_u}}$) is large.
	For a given ${{N_u}}$, the MCR-IM system surpasses the OrthoRa in terms of instantaneous throughput when ${\frac{{{E_{\rm{b}}}}}{{{N_0}}} > 5}$ dB, but performs worse when ${\frac{{{E_{\rm{b}}}}}{{{N_0}}} < 5}$ dB.
	When ${\frac{{{E_{\rm{b}}}}}{{{N_0}}} = 10}$ dB and ${{N_u} = 3}$, compared to OrthoRa, the throughput of the MCR-IM system with PD-SIC is ${21\% }$ and ${131\% }$ higher for ${{P_{frame}} = 2}$ and ${{P_{frame}} = 1}$, respectively.
	When ${{N_u} = 2}$, compared to the OrthoRa, the throughput of the MCR-IM system with PD-SIC is ${16\% }$ and ${129\% }$ higher for ${{P_{frame}} = 2}$ and ${{P_{frame}} = 1}$, respectively.
	It is worth noting that at ${\frac{{{E_{\rm{b}}}}}{{{N_0}}} = 10}$ dB, the throughput of the MCR-IM system is still better than that of the OrthoRa system, even when the direct demodulation algorithm is used.
	These results demonstrate that the MCR-IM system can distinguish the collision signals.
	Additionally, due to its higher transmission rate compared to other systems, the MCR-IM system requires fewer packet transmissions to send the same amount of data.
	As a result, the collision probability in the MCR-IM system is reduced, further enhancing the system capacity.
	
	\section{Conclusion}
	In this paper, we have proposed a MCR-IM system utilizing ZC sequences, which can significantly improve the transmission rate.
	Compared to traditional LoRa sequences, ZC sequences allow for a substantial increase in the number of parallel channels, making them well-suited for large-scale IoT scenarios.
	The MCR-IM system achieves quasi-orthogonality, thus allowing multiple users to transmit information simultaneously.	
	In addition, we have theoretically analyzed the BER performance of the MCR-IM system over Nakagami-${m}$ fading channel, and derived a closed-form BER expression.
	Simulation results show that compared with the FSCSS-IM system, the proposed system provides more than 0.4 dB SNR gain and carries more information bits.
	Compared to the GCSS system, MCR-IM yields an SNR gain of 0.2 dB, while transmitting the same number of bits.
	When ${L=4}$, the MCR-IM system with ${P=4}$ carries ${55\%}$ more bits than GCSS and ${34\%}$ more bits than FSCSS-IM.
	We have also compared the SE of the MCR-IM, FSCSS-IM, and GCSS systems.
	Results have shown that a larger ${P}$ value and a smaller ${L}$ value ensure that the MCR-IM system achieves a lower SNR performance for the same SE, meeting various IoT applications.
	Furthermore, we have proposed a PD-SIC algorithm to improve the collision decoding performance of the MCR-IM system.
	The computational complexity of the algorithm is proportional to the ${P}$ of the MCR-IM system.
	Simulation results have shown that when two colliding users are considerd, the BER of the MCR-IM system with the PD-SIC algorithm can reach ${10^{-5}}$, which is better than the OrthoRa system.
	Compared to the OrthoRa scheme (${{P_{frame}} = 2}$), the MCR-IM system using the PD-SIC algorithm can increase the throughput by ${16\%}$ and ${21\%}$ when there exist two and three colliding users, respectively.
	Owing to these advantages, the MCR-IM is well suited for the large-scale high-speed IoT applications.
	
	\appendices
	\section{}
	\label{proff LoRa}

	We consider two LoRa sequences ${{x_1}\left( n \right)}$ and ${{x_2}\left( n \right)}$ of equal length, denoted as
	\begin{align}
		{x_1}\left( n \right) = \frac{1}{{\sqrt N }}\exp \left( {j\frac{\pi }{N}\left( {{r_1}n + 2{q_1}} \right)n} \right),
	\end{align}
	\begin{align}
		{x_2}\left( n \right) = \frac{1}{{\sqrt N }}\exp \left( {j\frac{\pi }{N}\left( {{r_2}n + 2{q_2}} \right)n} \right),
	\end{align}
	where ${N = {2^{\nu}}}$. ${r}$ and ${q}$ are the chirp rate and the initial frequency, respectively.
	The cross-correlation between ${x_1}$ and ${x_2}$ is defined as
	\begin{align}
			&{\vartheta _1}\left( {{r_1},{r_2},{q_1},{q_2}} \right)\nonumber\\
			=& \frac{{\sum\limits_{n = 0}^{N - 1} {{x_1}\left( n \right)x_2^*\left( n \right)} }}{{\sqrt {\sum\limits_{n = 0}^{N - 1} {{{\left| {{x_1}\left( n \right)} \right|}^2}} } \sqrt {\sum\limits_{n = 0}^{N - 1} {{{\left| {{x_2}\left( n \right)} \right|}^2}} } }}\nonumber\\
			=& \frac{1}{N}\sum\limits_{n = 0}^{N - 1} {\exp \left( {j\frac{\pi }{N}\left( {\left( {{r_1} - {r_2}} \right)n + 2\left( {{q_1} - {q_2}} \right)} \right)n} \right)}.
	\end{align}

\begin{figure}[t]
		\centering
		\includegraphics[scale=0.36]{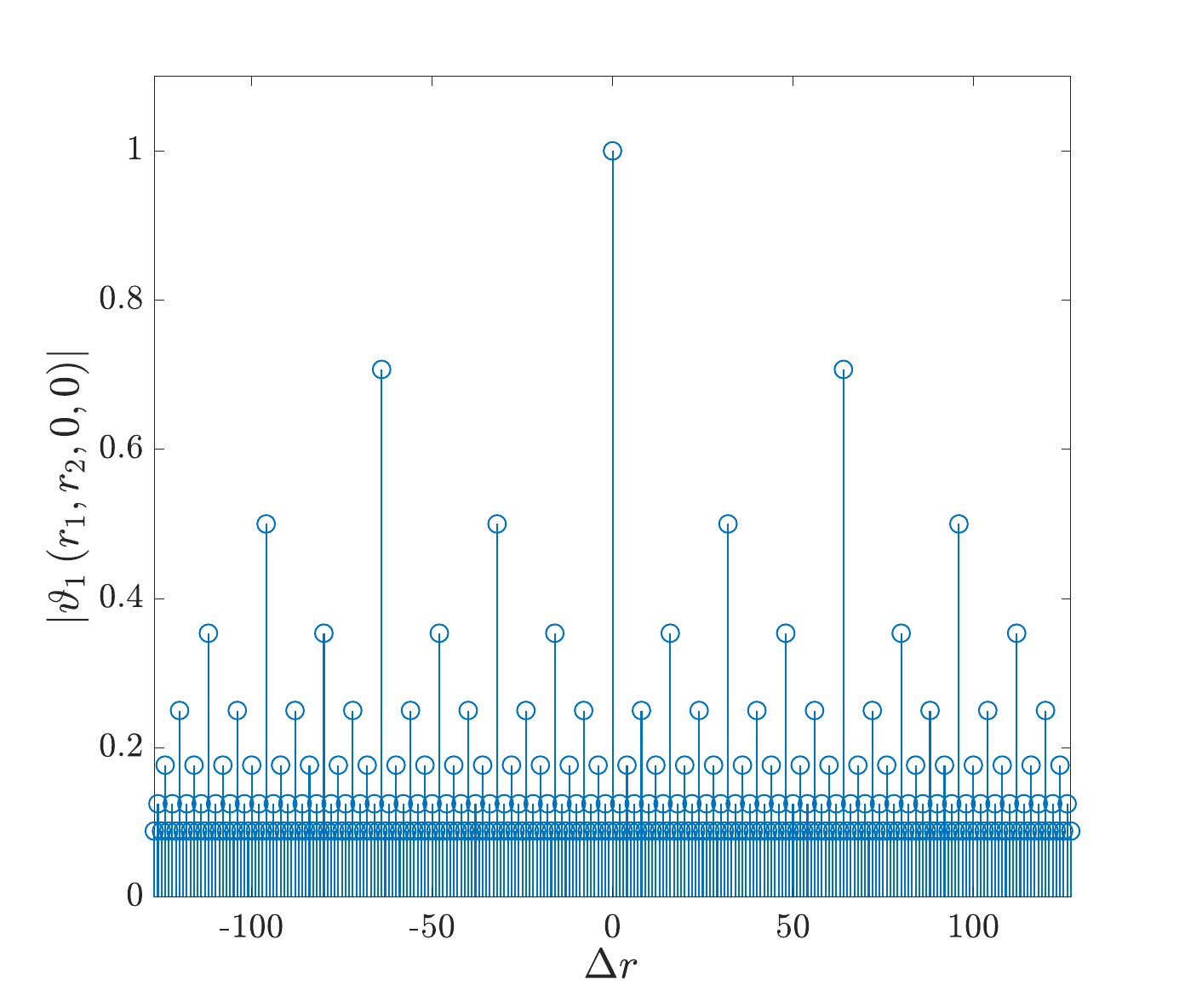}
		\caption{Cross-correlation of two conventional LoRa sequences at different chirp rates.}
		\label{fig10}
	\end{figure}

	We consider the cross-correlation between sequences with different chirp rates.
	We assume that ${{q_1} = {q_2} = 0}$, thus one has
	\begin{align}
		{\vartheta _1}\left( {{r_1},{r_2},0,0} \right) = \frac{1}{N}\sum\limits_{n = 0}^{N - 1} {\exp \left( {j\frac{\pi }{N}\Delta r{n^2}} \right)},
	\end{align}
	where ${\Delta r = {r_1} - {r_2}}$.
	Fig.~\ref{fig10} illustrates the relationship between ${\left| {{\vartheta _1}\left( {{r_1},{r_2},0,0} \right)} \right|}$ and ${\Delta r}$.
	Its observed that the cross-correlation between different LoRa sequences is relatively poor.
	The LoRa sequences is not suitable for multiple-chirp-rate index-modulation systems.
	
	\section{}
	\label{proff A}
	
	We consider two ZC sequences ${{x_1}\left( {N,{r_1},{q_1},n} \right)}$ and ${{x_2}\left( {N,{r_2},{q_2},n} \right)}$, whose cross-correlation is defined as
	\begin{align}
		&{\vartheta _1}\left( {{r_1},{r_2},{q_1},{q_2}} \right)\nonumber \\
		=& \frac{{\sum\limits_{n = 0}^{N - 1} {{x_1}\left( {N,{r_1},{q_1},n} \right)x_2^ * \left( {N,{r_2},{q_2},n} \right)} }}{{\sqrt {\sum\limits_{n = 0}^{N - 1} {{{\left| {{x_1}\left( {N,{r_1},{q_1},n} \right)} \right|}^2}} } \sqrt {\sum\limits_{n = 0}^{N - 1} {{{\left| {{x_2}\left( {N,{r_2},{q_2},n} \right)} \right|}^2}} } }}\nonumber\\
		=& \sum\limits_{n = 0}^{N - 1} {{x_1}\left( {N,{r_1},{q_1},n} \right)x_2^ * \left( {N,{r_2},{q_2},n} \right)}.
	\end{align}

	It is straightforward to show that ${{\vartheta _1}\left( {{r_1},{r_2},{q_1},{q_2}} \right) = 1}$ when ${{r_1} = {r_2}}$ and ${{q_1} = {q_2}}$.
	This represents the point where cross-correlation between the two sequences is strongest.
	When ${{r_1} = {r_2}}$ and ${{q_1} \ne {q_2}}$, one can obtain
	\begin{align}
 		{\vartheta _1}\left( {{r_1},{r_2},{q_1},{q_2}} \right) = \frac{1}{N}\sum\limits_{n = 0}^{N - 1} {\exp \left( {j\frac{{2\pi }}{N}cn} \right)}  = 0.
	\end{align}

	This is due to the fact that ${c}$ and ${N}$ are coprime, where ${c = {r_1}\left( {{q_1} - {q_2}} \right)}$.
	When ${{r_1} \ne {r_2}}$, one can obtain
	\begin{align}
		&\left| {{\vartheta _1}\left( {{r_1},{r_2},{q_1},{q_2}} \right)} \right|\nonumber\\
		=& \frac{1}{N}\left| {\sum\limits_{n = 0}^{N - 1} {\exp \left( {j\frac{\pi }{N}{r_3}\left( {n + 1 + 2{q_3}} \right)n} \right)} } \right|\nonumber\\
		=& \frac{1}{{\sqrt N }},
	\end{align}
	where ${{r_3} = {r_1} - {r_2}}$ and ${{q_3} = \frac{{{q_1}{r_1} - {q_2}{r_2}}}{{{r_3}}}}$.
	The detailed derivation process can be found in \cite{144727}.
	In summary, the ZC sequences with different chirp rates exhibit a quasi-orthogonal relationship.
	
	\section{}
	\label{proff C}
	
	In traditional LoRa systems, signals of different SFs exhibit quasi-orthogonal properties.
	We expect this advantageous property to remain valid in MCR-IM systems.
	We consider two different SF signals ${{X_1}\left( n \right)}$ and ${{X_2}\left( n \right)}$ with lengths ${N_1}$ and ${N_2}$, respectively, where ${{N_1} > {N_2}}$, ${{X_2}\left( n \right)}$ contains a random time delay ${m_d}$ satisfying ${{m_d} \in \left\{ {0,1, \cdots ,{N_1} - {N_2}} \right\}}$. The two signals can be described as
	\begin{align}
		{X_1}\left( n \right) = \sum\limits_{i = 0}^{L - 1} {{x_{1,i}}\left( {{N_1},{r_{1,i}},{q_{1,i}},n} \right)},
	\end{align}
	\begin{align}
		{X_2}\left( n \right) = \sum\limits_{j = 0}^{L - 1} {{x_{2,j}}\left( {{N_2},{r_{2,j}},{q_{2,j}},n - {m_d}} \right)} {G_{{m_d},{N_2} + {m_d} - 1}}\left( n \right),
	\end{align}
	where ${n = 0,1, \cdots {N_1} - 1}$ and
	\begin{align}
		\label{Gab}
		{G_{a,b}}\left( n \right) = \left\{ {\begin{array}{*{20}{c}}
				{1,\quad a \le n \le b,}\\
				{0,\qquad \quad {\rm{other}}.}
		\end{array}} \right.
	\end{align}

	The cross-correlation between ${{X_1}\left( n \right)}$ and ${{X_2}\left( n \right)}$ is defined as
	\begin{align}
		{\vartheta _2} &= \frac{{\sum\limits_{n = 0}^{{N_1} - 1} {X_1^ * \left( n \right){X_2}\left( n \right)} }}{{\sqrt {\sum\limits_{n = 0}^{{N_1} - 1} {{{\left| {{X_1}\left( n \right)} \right|}^2}} } \sqrt {\sum\limits_{n = 0}^{{N_1} - 1} {{{\left| {{X_2}\left( n \right)} \right|}^2}} } }}\nonumber\\
		&= \frac{1}{{\sqrt {{N_1}{N_2}} }}\sum\limits_{n = {m_d}}^{{N_2} + {m_d} - 1} {\sum\limits_{i = 0}^{L - 1} {x_{1,i}^ * \left( n \right)} \sum\limits_{j = 0}^{L - 1} {{x_{2,j}}\left( {n - {m_d}} \right)} }.
	\end{align}

	From Appendix \ref{proff B}, one has
	\begin{align}
		\left| {\frac{1}{{\sqrt {{N_1}{N_2}} }}\sum\limits_{n = {m_d}}^{{N_2} + {m_d} - 1} {x_{1,i}^ * \left( n \right){x_{2,j}}\left( {n - {m_d}} \right)} } \right| \le \sqrt {\frac{{1 + 2\varepsilon }}{{{N_1}}}},
	\end{align}
	where ${\varepsilon }$ is a relatively small number.
	One can get
	\begin{align}
		\left| {{\vartheta _2}} \right| \le {L^2}\sqrt {\frac{{1 + 2\varepsilon }}{{{N_1}}}}.
	\end{align}

	Due to the cancellation of the real and imaginary components, we can assume that transmitted signals with different SFs exhibit quasi-orthogonal relationship.
	This assumption holds because the actual cross-correlation value is significantly lower than this upper bound and ${N_1}$ is typically large.
	
	\section{}
	\label{proff B}
	
	We consider two ZC sequences ${{x_1}\left( {{N_1},{r_1},{q_1},n} \right)}$ and ${{x_2}\left( {{N_2},{r_2},{q_2},n} \right)}$, corresponding to different SFs with effective lengths ${N_1}$ and ${N_2}$, respectively.
	Moreover, ${{x_2}\left( {{N_2},{r_2},{q_2},n} \right)}$ involves an arbitrary delay ${{m_d}}$ satisfying ${{m_d} \in \left\{ {0,1, \cdots ,{N_1} - {N_2}} \right\}}$.
	They can be expressed as
	\begin{align}
		{x_1}\left( {{N_1},{r_1},{q_1},n} \right) = {e^{\left( {j\frac{\pi }{{{N_1}}}{r_1}\left( {n + 1 + 2{q_1}} \right)n} \right)}},
	\end{align}
	\begin{align}
		&{x_2}\left( {{N_2},{r_2},{q_2},n - {m_d}} \right)\nonumber\\
		=& {\Lambda _1}{e^{\left( {j\frac{{\pi {r_2}\left( {n - 2{m_d} + 1 + 2{q_2}} \right)n}}{{{N_2}}}} \right)}}{G_{{m_d},{N_2} + {m_d} - 1}}\left( n \right),
	\end{align}
	\begin{align}
		{\Lambda _1} = {e^{j\frac{\pi }{{{N_2}}}{r_2}\left( {m_d^2 - {m_d} - 2{m_d}{q_2}} \right)}},
	\end{align}
	where ${{G_{a,b}}\left( n \right)}$ is defined in (\ref{Gab}) and ${n = 0,1, \cdots {N_1} - 1}$.
	Here, ${{\Lambda _1}}$ represents a phase shift introduced in ${{x_2}\left( {{N_2},{r_2},{q_2},n - {m_d}} \right)}$, which is induced by the delay ${m_d}$.
	The cross-correlation between ${{x_1}\left( {{N_1},{r_1},{q_1},n} \right)}$ and ${{x_2}\left( {{N_2},{r_2},{q_2},n - {m_d}} \right)}$ is defined as
	\begin{align}
		&{\vartheta _2}\left( {{N_1},{N_2},{r_1},{r_2},{q_1},{q_2},{m_d}} \right)\nonumber\\
		=& \frac{{\sum\limits_{n = 0}^{{N_1} - 1} {x_1^ * \left( {{N_1},{r_1},{q_1},n} \right){x_2}\left( {{N_2},{r_2},{q_2},n - {m_d}} \right)} }}{{\sqrt {\sum\limits_{n = 0}^{{N_1} - 1} {{{\left| {{x_1}\left( {{N_1},{r_1},{q_1},n} \right)} \right|}^2}} } \sqrt {\sum\limits_{n = 0}^{{N_1} - 1} {{{\left| {{x_2}\left( {{N_2},{r_2},{q_2},n - {m_d}} \right)} \right|}^2}} } }}\nonumber\\
		=& \frac{1}{{\sqrt {{N_1}{N_2}} }}\sum\limits_{n = {m_d}}^{{N_2} + {m_d} - 1} {x_1^ * \left( {{N_1},{r_1},{q_1},n} \right){x_2}\left( {{N_2},{r_2},{q_2},n - {m_d}} \right)}.
	\end{align}

	After simplification, one obtains
	\begin{align}
		\label{correlation 1}
		&{\vartheta _2}\left( {{N_1},{N_2},{r_1},{r_2},{q_1},{q_2},{m_d}} \right)\nonumber\\
		=& \frac{{{\Lambda _1}{\Lambda _2}}}{{\sqrt {{N_1}{N_2}} }}\sum\limits_{n = 0}^{{N_2} - 1} {{e^{j2\pi \left( {a{n^2} + bn} \right)}}},
	\end{align}
	\begin{align}
		{\Lambda _2} = {e^{j\pi {m_d}\left( {\frac{{{r_2}\left( {1 - {m_d} + 2{q_2}} \right)}}{{{N_2}}} - \frac{{{r_1}\left( {1 + {m_d} + 2{q_1}} \right)}}{{{N_1}}}} \right)}},
	\end{align}
	where ${a = \frac{{{r_2}}}{{2{N_2}}} - \frac{{{r_1}}}{{2{N_1}}}}$ and ${b = \frac{{{r_2}\left( {1 + 2{q_2}} \right)}}{{2{N_2}}} - \frac{{{r_1}\left( {1 + 2{m_d} + 2{q_1}} \right)}}{{2{N_1}}}}$.
	
	According to Lemma 2 of Appendix B in \cite{10045787}, (\ref{correlation 1}) can be further formulated as
	\begin{align}
		{\vartheta _2}\left( {{N_1},{N_2},{r_1},{r_2},{q_1},{q_2},{m_d}} \right) = \frac{{{\Lambda _1}{\Lambda _2}}}{{\sqrt {{N_1}{N_2}} }}\lambda {e^{j\theta }},
	\end{align}
	where
	\begin{align}
		\theta  = {\tan ^{ - 1}}\frac{{\sum\limits_{n = 0}^{{N_2} - 1} {\sin \left( {2\pi \left( {a{n^2} + bn} \right)} \right)} }}{{\sum\limits_{n = 0}^{{N_2} - 1} {\cos \left( {2\pi \left( {a{n^2} + bn} \right)} \right)} }},
	\end{align}
	\begin{align}
		\lambda  = \sqrt {{N_2} + 2\sum\limits_{n = 0}^{{N_2} - 1} {\sum\limits_{i = n + 1}^{{N_2} - 1} {\cos \left( {2\pi \left( {n - i} \right)\left( {a\left( {n + i} \right) + b} \right)} \right)} } }.
	\end{align}

	It can be straightforwardly shown that ${\left| {{\vartheta _2}\left( {{N_1},{N_2},{r_1},{r_2},{q_1},{q_2},{m_d}} \right)} \right|}$ is upper bounded by
	\begin{align}
		\left| {{\vartheta _2}\left( {{N_1},{N_2},{r_1},{r_2},{q_1},{q_2},{m_d}} \right)} \right| \le \sqrt {\frac{{1 + 2\varepsilon }}{{{N_1}}}} ,
	\end{align}
	where
	\begin{align}
		\varepsilon  = \max \left( {\sum\limits_{i = n + 1}^{{N_2} - 1} {\cos \left( {2\pi \left( {n - i} \right)\left( {a\left( {n + i} \right) + b} \right)} \right)} } \right).
	\end{align}
	
	One can see that the magnitude of the cross-correlation between the ${{x_1}\left( {{N_1},{r_1},{q_1},n} \right)}$ and ${{x_2}\left( {{N_2},{r_2},{q_2},n} \right)}$ with different SFs is inversely proportional to ${N_1}$, which corresponds to a larger SF.
	Therefore, ZC sequences with different SFs can be considered as quasi-orthogonal.

	\bibliographystyle{IEEEtr}
	\bibliography{reference}
	
\end{document}